\documentclass[prb,amsfonts,amssymb,floats,twocolumn,superscriptaddress,aps]{revtex4-2}
\usepackage[utf8]{inputenc}
\usepackage[T1]{fontenc}
\usepackage{amsmath}
\usepackage{float}
\usepackage[caption = false]{subfig}
\usepackage[dvipsnames]{xcolor}
\usepackage{braket}
\usepackage{bm}
\usepackage{dsfont}
\usepackage{adjustbox}
\usepackage{graphicx}
\usepackage{enumitem}
\usepackage{bbm}
\usepackage{tikz}
\usepackage{physics}
\usepackage{microtype}

\usepackage{txfonts}

\usepackage[unicode=true,colorlinks=true,pdfpagemode=UseOutlines,pdfstartview=Fith]{hyperref}
\hypersetup{linkcolor=blue,citecolor=blue,urlcolor=blue}

\usepackage{slashed}
\usepackage{orcidlink}

\makeatletter

\def\QED3{$\text{QED}_3$}

\renewcommand{\Re}{\operatorname{Re}} 
\renewcommand{\Im}{\operatorname{Im}} 

\newcommand{\iu}{\mathrm{i}} 
\newcommand{\eu}{\mathrm{e}} 
\newcommand{\du}{\mathrm{d}} 
\newcommand{\hc}{\mathrm{h.c.}} 

\newcommand{\bvec}[1]{{\bm{#1}}}

\newcommand{\Ztwo}{\mathbb{Z}_2}
\newcommand{\Uone}{\mathrm{U(1)}}
\newcommand{\SUtwo}{\mathrm{SU(2)}}
\newcommand{\SUfour}{\mathrm{SU(4)}}

\newcommand{\SOthree}{\mathrm{SO(3)}}

\newcommand{\SO}{\mathrm{SO}}

\makeatother	

\begin{document}

	\title{Spin-Peierls instability of the U(1) Dirac spin liquid}
	
	\author{Urban F. P. Seifert}
\thanks{These two authors contributed equally to this work.}
\affiliation{Kavli Institute for Theoretical Physics, University of California, Santa Barbara, CA 93106}

 \author{Josef Willsher \orcidlink{0000-0002-6895-6039}}
\thanks{These two authors contributed equally to this work.}
\affiliation{Technical University of Munich, TUM School of Natural Sciences, Physics Department, 85748 Garching, Germany}
\affiliation{Munich Center for Quantum Science and Technology (MCQST), Schellingstr. 4, 80799 M{\"u}nchen, Germany}

\author{Markus Drescher \orcidlink{0000-0001-9231-1882}}
\affiliation{Technical University of Munich, TUM School of Natural Sciences, Physics Department, 85748 Garching, Germany}
\affiliation{Munich Center for Quantum Science and Technology (MCQST), Schellingstr. 4, 80799 M{\"u}nchen, Germany}

\author{Frank Pollmann \orcidlink{0000-0003-0320-9304}}
\affiliation{Technical University of Munich, TUM School of Natural Sciences, Physics Department, 85748 Garching, Germany}
\affiliation{Munich Center for Quantum Science and Technology (MCQST), Schellingstr. 4, 80799 M{\"u}nchen, Germany}

 	\author{Johannes Knolle \orcidlink{0000-0002-0956-2419}}
\affiliation{Technical University of Munich, TUM School of Natural Sciences, Physics Department, 85748 Garching, Germany}
\affiliation{Munich Center for Quantum Science and Technology (MCQST), Schellingstr. 4, 80799 M{\"u}nchen, Germany}
\affiliation{\small Blackett Laboratory, Imperial College London, London SW7 2AZ, United Kingdom}

	\begin{abstract}
A complicating factor in the realization and observation of quantum spin liquids in materials is the ubiquitous presence of other degrees of freedom, in particular lattice distortion modes (phonons). These provide additional routes for relieving magnetic frustration, thereby possibly destabilizing spin-liquid ground states.
In this work, we focus on triangular-lattice Heisenberg antiferromagnets, where recent numerical evidence suggests the presence of an extended U(1) Dirac spin liquid phase which is described by compact quantum electrodynamics in 2+1 dimensions (QED$_3$), featuring gapless spinons and monopoles as gauge excitations, and believed to flow to a strongly-coupled fixed point with conformal symmetry. Using complementary perturbation theory and scaling arguments, we show that a symmetry-allowed coupling between (classical) finite-wavevector lattice distortions and monopole operators of the U(1) Dirac spin liquid generally induces a spin-Peierls instability towards a (confining) 12-site valence-bond solid state.
We support our theoretical analysis with state-of-the-art density matrix renormalization group simulations.
Away from the limit of static distortions, we demonstrate that the phonon energy gap establishes a parameter regime where the spin liquid is expected to be stable, and show that the monopole-lattice coupling leads to softening of the phonon in analogy to the Kohn anomaly.
We discuss the applicability of our results to similar systems, in particular the Dirac spin liquid on the Kagome lattice.
\end{abstract}
	\maketitle

\section{Introduction}
The presence of many competing classical ground states in frustrated magnets implies that strong quantum fluctuations may stabilize quantum spin liquids (QSL) as highly exotic states of \emph{quantum matter}. These states fall outside Landau's paradigm of symmetry-breaking orders, and are instead characterized by a rich entanglement structure and described in terms of emergent deconfined gauge theories \cite{wen2004quantum,savbalents16,doi:10.1146/annurev-conmatphys-031218-013401}.

Recent years have witnessed remarkable progress in the theoretical description and classification of QSLs, as well as the experimental identification and characterization of candidate materials.
However, while model spin systems have been studied using a variety of theoretical and numerical methods, definitive experimental evidence for the realization of a QSL phase in a material is still outstanding.
Materials inevitably include additional spin-spin interactions, impurities and other sources of quenched disorder \cite{willans10,zhu17,PhysRevX.8.031028,knolle2019bond}, as well as the coupling of spins to additional degrees of freedom, for example lattice distortions.
Given the competing nature of interactions in frustrated magnets \cite{mila11}, these perturbations can have a marked impact \cite{PhysRevLett.106.067203,sybalents16,PhysRevB.97.134432} on characteristic observable properties of QSLs realized in actual materials, or even mask such a phase completely, resulting in possibly complex phase diagrams of multiple competing orders~\cite{rau2014generic,cookmeyer2021four}.

A paradigmatic example for the interplay of strong spin fluctuations and lattice degrees of freedom is given by the spin-Peierls instability of one-dimensional spin-1/2 chains:
While the canonical antiferromagnetic Heisenberg chain possesses a gapless disordered ground state with power-law decaying correlations, an infinitesimally weak coupling to static lattice deformations leads to a finite dimerization and the concomitant opening of a spin gap \cite{pincus71,pytte74}, since the energy gain of dimerization of the spin liquid outcompetes the harmonic elastic energy cost.
This presupposes that the coupling of lattice distortion to the dimerization operator (instanton) of the effective Luttinger-liquid field theory for the $S=1/2$ Heisenberg chain is \emph{symmetry-allowed} and \emph{relevant}.
Accounting for intrinsic quantum dynamics of the lattice degrees of freedom (i.e. phonons with finite frequency), a \emph{finite} interaction strength is required to induce the ordering transition \cite{orignac04,giamarchi2004quantum}.

Turning to two-dimensional frustrated magnets, experiments show that magnetic phase transitions and the nature of magnetic ordering are often correlated with structural distortions which can act to (partially) relieve magnetic frustration.
A crucial, long-standing question of immense experimental relevance concerns the \emph{stability} of spin-liquid states against spin-lattice couplings.
While numerical studies have found evidence for the presence of spin-Peierls-type instabilities in gapless spin-disordered phases \cite{becca02,ferrari21}, there are only few analytical results for \emph{gapless spin-liquids}, which have been mostly focused on intrinsic nesting instabilities of spinon Fermi surfaces \cite{PhysRevLett.115.177205,PhysRevB.104.165133}. 
It is well understood that frustrated spin-systems may form valence bond solid (VBS) states via spontaneous symmetry breaking \cite{PhysRevLett.62.1694,PhysRevB.42.4568}.
However, a destabilization of a stable spin liquid ground state by dynamically generating lattice distortions, the true analog to the one-dimensional instability, has hereto not been investigated.

In the manuscript at hand, we consider the U(1) Dirac spin liquid (DSL) and make use of recent numerical and field-theoretic advances to investigate its stability against spin-lattice couplings.
The DSL is a paradigmatic example of a QSL characterized by strongly interacting gapless spinons and U(1) gauge fluctuations which are described (at low energies) by compact quantum electrodynamics in 2+1 dimensions (QED$_3$).
Initially suggested as a candidate ground state for square-lattice cuprates \cite{lee06}, the DSL later gained support as describing the spin-disordered phase of frustrated Heisenberg antiferromagnets on kagome \cite{ran07,herm08,he17,iqbal15,budaraju2023piercing} and triangular lattices \cite{PhysRevB.42.4800,iqbal16,PhysRevLett.123.207203,drescher2022dynamical}.

At low energies, QED$_3$ is believed to flow to a strongly coupled fixed point with conformal symmetry, allowing for the low-energy description of the U(1) DSL and its excitations in terms of a conformal field theory (CFT)~\cite{appel86, karthik16,chester16,alba22}.
Crucially, for the DSL to be an \emph{intrinsically} stable phase of matter, all relevant operators in this conformal field theory must transform non-trivially under the \emph{microscopic} symmetries of the system, such that they cannot be generated during the symmetry-preserving flow from the UV to the IR.

The most relevant operators in the QED$_3$ field theory are monopoles which can be understood as topological instanton events which tunnel an integer-$2 \pi$ magnetic flux of the emergent $\Uone$ gauge field.
Proliferation of such monopoles corresponds to a confinement transition.
As shown by Song \textit{et al.} in Refs.~\cite{song19,song20} and found in numerical simulations \cite{drescher2022dynamical,wietek23}, monopole operators transform under microscopic symmetries and can be identified with order parameters for Néel-ordered antiferromagnetic and VBS phases, which can hence be accessed via a proliferation of these monopole operators --- the U(1) DSL state has therefore also been dubbed the ``mother of competing orders'' \cite{herm05}.
Importantly, the proliferation of monopoles on the triangular and kagome lattices is forbidden as long as the \emph{microscopic} lattice symmetries are preserved, and on these lattices the DSL is expected to be stable.

The Luttinger liquid phase of 1+1 dimensional Heisenberg spin-half chains can also be said to be the mother of competing VBS and AFM orders \cite{PhysRevB.99.165143}.
Similarly it has an equivalent description in terms of Dirac fermions coupled to a dynamical U(1) gauge field \cite{Kim:1999aa,PhysRevB.49.5200,Hosotani_1997} and therefore may be understood as a 1+1 dimensional analog to the DSL \cite{song19}.
This analogy between theories will be our basis for extending the well-established spin-Peierls instability of spin chains to 2+1 dimensional QSLs.

In this work, we show that on the triangular lattice, there exists a symmetry-allowed coupling between monopoles and lattice deformations at certain (finite) wavevectors.
Though the monopoles \emph{per se} are relevant operators in the QED$_3$ CFT and the coupling is symmetry-allowed, this does not \emph{necessarily} imply an instability of the DSL, since there is an intrinsic (elastic) energy cost associated with a lattice distortion.
For an instability coupling to occur (at nonzero coupling), the system must \emph{dynamically} generate such a distortion by optimizing the competition between the elastic potential and the possible distortion-induced energy gain of the spin liquid.

Here, we exploit the conformal nature of the QED$_3$ fixed point theory \cite{alba22} and use both weak-coupling conformal perturbation theory as well as scaling approaches to analyze the coupled lattice-DSL system, and show that at zero temperature, any infinitesimally weak coupling $g$ induces a finite distortion  of the underlying lattice and simultaneously precipitates monopole proliferation, leading the VBS order.

\begin{figure}
\centering
\includegraphics[width=\columnwidth]{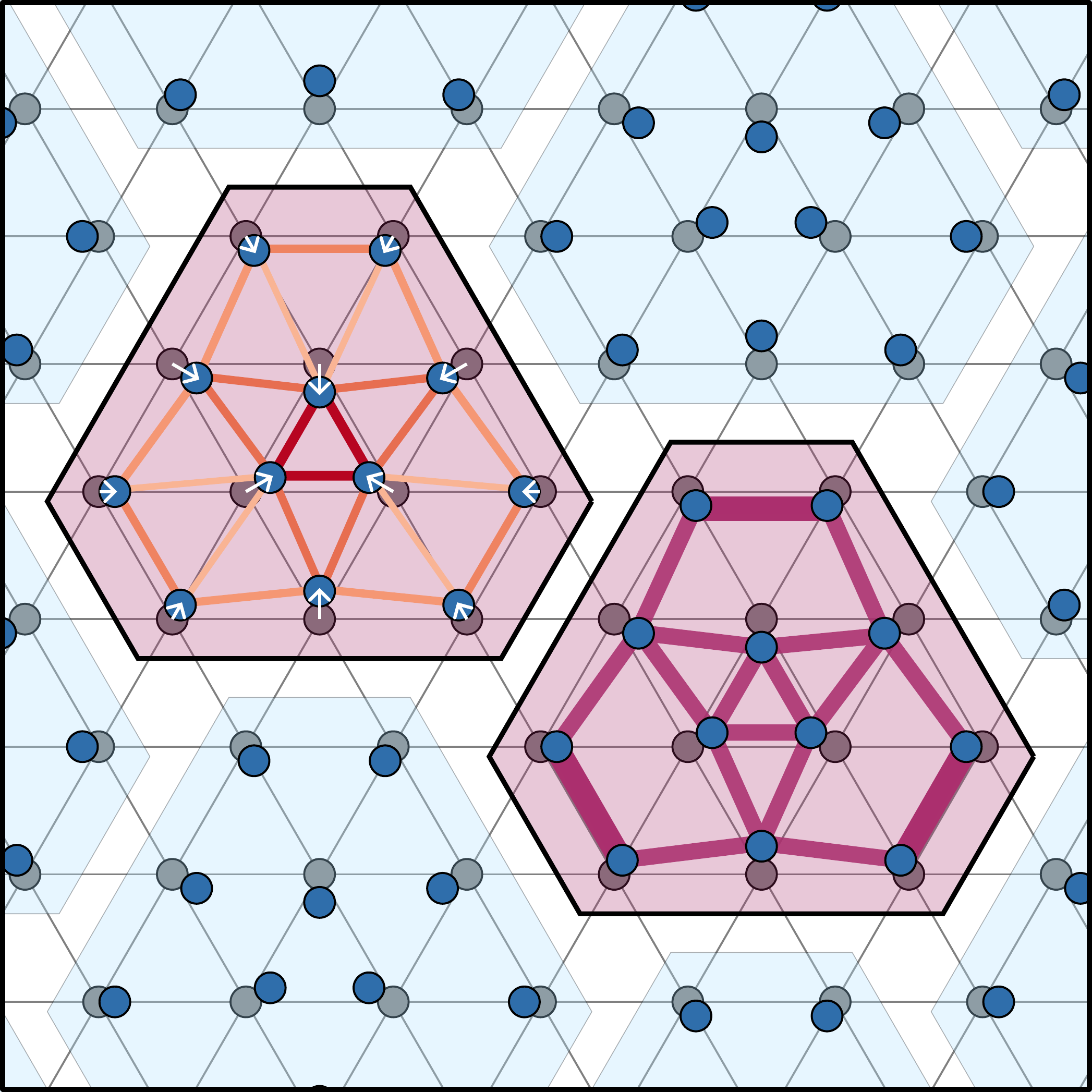}
\caption{Due to strong interactions of gapless gauge excitations in the U(1) DSL on the triangular lattice, it spontaneously distorts and forms 12-site VBS plaquette pattern (coloured). The blue shifted dots show the distorted lattice, the vector displacements inside the unit cell $\bvec{u}(\bvec{x})$ are highlighted by white arrows. The enhanced nearest-neighbor bond strengths within each plaquette are depicted in the left panel, and the short-range spin-spin correlations are shown on the right.}
\label{fig1}
\end{figure}

To make connection with experiment, we establish the phenomenology of the spin-Peierels instability of DSL and explicitly determine the symmetry of the lattice distortion and corresponding 12-site VBS-ordered state, as depicted in  Figure~\ref{fig1}.
Our field-theoretic approach further allows us to determine critical temperatures for the ordering transition, which exhibit a non-trivial power-law dependence on the spin-lattice coupling $T_{\mathrm{SP}} \sim g^{1/(3-2\Delta_\Phi)}$ (with $\Delta_\Phi$ denoting the scaling dimension of the monopole operator). 

Guided by these theoretical predictions, we make use of state-of-the-art density matrix renormalization group (DMRG) \cite{White1992, White1993, McCulloch2008} methods to numerically simulate the antiferromagnetic extended Heisenberg model on a distorted triangular lattice \cite{PhysRevB.42.4800,PhysRevLett.123.207203, Sherman2022, drescher2022dynamical}.
We stabilize the DSL state on the undistorted geometry, and compare the energy gain of several lattice distortion patterns; we find that the largest energy gain is observed for the 12-site lattice distortion corresponding to the spin-Peierls transition from monopole condensation.
We use finite-size scaling to argue that this is consistent with a zero-coupling instability in the thermodynamic limit as predicted by the field theory analysis.

Beyond the adiabatic approximation, where lattice deformations are considered static, finite phonon frequencies $\omega_0$ induce retarded monopole-monopole interactions.
An analysis of the resulting effective action shows that there exists a finite parameter regime $g^2 < c \omega_0^{3-2\Delta_\Phi}$ where the DSL is expected to be stable (for some constant $c$).
Our work, therefore, places constraints on the experimental observability of U(1) Dirac spin liquids in systems with non-negligible spin-lattice couplings. As an additional experimental signature of the DSL in the stable regime we point out that the DSL-lattice coupling leads to a Kohn-type anomaly of the phonon spectrum.

The manuscript has the following structure.
In Section~\ref{sec:spinpeierls}, we briefly review the Peierls instability in both free and interacting one-dimensional systems and discuss previous approaches to two-dimensional Peierls phases.
Sec.~\ref{sec:DSL} introduces the $\Uone$ Dirac spin liquid ground state of the frustrated triangular-lattice Heisenberg model and its conformal field theory description. 
In Sec.~\ref{sec:coupling} we derive symmetry-allowed couplings between monopole operators in the effective theory of the DSL and lattice distortions.
We investigate the spin-Peierls instability in Sec.~\ref{sec:instability} using complementary weak-coupling perturbation theoretic and scaling arguments.
In Sec.~\ref{sec:numerics} we supplement our analytical insights with numerical DMRG results.
Finally, in Sec~\ref{sec:dyn-phon} we discuss the case of quantum (finite-frequency) phonons. We argue for a finite parameter window of stability and show that a temperature-dependent softening of the phonon dispersion provides a potentially accessible signature of the DSL.
Sec.~\ref{sec:conclusion} concludes the paper and discusses the relevance of our results for recent triangular- and kagome-lattice QSL candidate materials.

\section{Recap of the spin-Peierls instability}\label{sec:spinpeierls}

\subsection{Free spins \& fermions}
Peierls' theorem can be succinctly summarized as stating that a chain of equally spaced ions with odd numbers of free fermions per site is unstable for any infinitesimal coupling to the lattice \cite{peierls1996quantum}. 
Considering a two-site lattice distortion $u_i = (-1)^i \delta u$, the potential energy cost in the Harmonic approximation of the ions will be $H_{\mathrm{elast}} = (K/2) \sum_{i} (u_i-u_{i+1})^2$, with lattice stiffness $K$. This distortion has lattice wavevector $k=\pi/2$ and is commensurate with the Fermi wavevector for a half-filled band of fermions.
This lattice deformation correspondingly changes the fermion tight-binding hopping amplitudes at first order to $t_i = t + (-1)^i \delta t$ for $\delta t / t = g(\delta u / \mathsf{a})$, where $\mathsf{a}$ is the lattice spacing, and $g$ is a dimensionless constant which quantifies the change in overlap. The spectrum opens a band gap $\delta t$ and the electronic energy of the lower occupied band is reduced; the total energy change is
\begin{equation}
    E(\delta u)_{\mathrm{free}} = E_{\mathrm{GS}} + (\delta u )^2 \left [K - \frac{t\, g^2}{2 \mathsf{a}^2} \log(\frac{g^2 \, (\delta u)^2 }{\mathsf{a}^2}) \right].
\end{equation}
Due to the appearance of the logarithm in this approximation, the energy gain out-competes the cost of static lattice distortion and a finite equilibrium dimerization is developed $\delta u_{\mathrm{eq}} \sim \mathsf{a} \exp[-{K \mathsf{a}^2}/{(g^2 t)}]$, giving alternating strong and weak hopping amplitudes as seen in polyacetylene \cite{PhysRevLett.42.1698,RevModPhys.60.781}.
The XY chain can be exactly mapped via the Jordan--Wigner transformation to free spinless one-dimensional fermions. The model's gapless, spin-disordered ground state is destroyed by the spin-Peierls dimerization of the lattice at infinitesimal coupling \cite{pincus71,pytte74,FUKUYAMA198763}.
The resulting valence-bond order of the spins consists of singlet states on each strong bond.

One can attempt to generalize the one-dimensional result by coupling gapless Dirac fermions in half-filled graphene to commensurate lattice distortions at the $\bvec{K}$-points.
Indeed, the proposed Kekul\'e distortion of the underlying honeycomb lattice modulates the hopping amplitudes in exactly this way and opens up a gap in the Dirac spectrum \cite{PhysRevLett.98.186809,PhysRevB.77.235431}.
However, the electronic energy gain from the resulting gapped spectrum in 2D is not logarithmically divergent as in 1D, but finite because of the vanishing fermionic density of states at the Dirac points.
It leads to a finite fermion-lattice coupling $g_c = [2\mathsf{a}^2 K / 3t]^{1/2} > 0$, below which there is no dimerisation in the ground state $\delta u_{\mathrm{eq}} = 0$. 
Due to the absence of a weak-coupling Peierls instability, pure graphene does not show this Kekul\'e distortion (unless in the form of quasi-one-dimensional carbon nanotube \cite{PhysRevB.62.2806}), but only engineered materials with sufficient interactions can induce it, for example using an atom-trap platform \cite{PhysRevLett.98.186809}, surface disorder \cite{doi:10.1126/sciadv.abm5180}, or lithium ion intercalation \cite{Gutierrez:2016aa,PhysRevLett.126.206804}. In the context of QSLs, recent work has shown that the emergent, free, gapless Majorana fermions of the Z$_2$ Kitaev honeycomb-lattice QSL can be gapped by the same Kekul\'e distortion above a critical coupling to the lattice $g_c>0$ \cite{PhysRevB.99.155138,PhysRevB.101.245116}.

\subsection{Interacting spins in one dimension}\label{sec:luttsp}

We now move on to spin-Peierls instabilities of interacting systems, focusing on the 1D antiferromagnetic ($J>0$) Heisenberg chain with the Hamiltonian 
\begin{equation}
    H_{\mathrm{Heis}} =  J \sum_i \vec{S}_{i}\cdot\vec{S}_{i+1},
\end{equation} 
with $\SUtwo$ spin operators $S^\alpha$, $\alpha = x,y,z$.
It can be rewritten in terms of interacting spinless fermions via a Jordan--Wigner transformation. Within the bosonization framework \cite{FUKUYAMA198763,giamarchi2004quantum} it can be shown that the lowest-energy excitations correspond to fluctuations in the fermion density $\partial_x \phi(x)$ and phase. 
The resulting field theoretic Hamiltonian $H[\phi(x)]$, where $\phi(x) \in [0,2\pi]$, takes the form of the Luttinger liquid:
\begin{multline}\label{eq:ll}
  H[\phi(x)] = \frac{1}{\pi} \int\dd[2]{x} \left[ \frac{1}{4} (\partial_t \phi)^2 +  (\partial_x \phi)^2 \right]
  + V_4 \int\dd[2]{x} \cos(4\phi).
\end{multline}
The parameter $V_4$ at the UV scale is fixed by the lattice parameter $J$.
The kinetic term describes a Luttinger liquid, a free boson with fixed normalization $1/2$ (see Appendix~\ref{app:bozonization} for the general bosonization description of spin chains in terms of Luttinger liquids).
The $V_4$ coupling to the cosine represents an instanton term, which is marginally irrelevant, meaning the Heisenberg model has a gapless ground state.

The standard coupling of the spins to the lattice \cite{PhysRevB.70.214436} is then given by 
\begin{equation}
 H_g = g \sum_i  (u_i - u_{i+1}) \vec{S}_{i}\cdot\vec{S}_{i+1},
\end{equation}
and is quantified by the coupling $g$. The total Hamiltonian is given as the sum $H = H_\mathrm{Heis} + H_\mathrm{elast} + H_g$.
Bosonizing this interaction leads to a new instanton term in the effective theory
\begin{equation}
    H[u] = V_2 \int \dd[2]{x} \sin(2\phi),
    \quad\text{with}\,\,
    V_2 =g\,\delta u,
\end{equation}
where the coupling $V_2$ in the Heisenberg model has dimension $\Delta_2 = [V_2] = 1/2$ which is more relevant than $\Delta_4$.
Note that translating by one lattice constant $i\to i+1$, we have $\sin(2\phi) \to -\sin(2\phi)$ and $\delta u \to - \delta u$ and the action still obeys translational symmetry by construction.

The mass scale generated by the coupling $V_2 = g\, \delta u$ is given by $(g\,\delta u)^{\chi/2}$ and the gapped ground state energy is lowered by $(g\,\delta u)^{\chi}$ where $\chi=4/3$ for the AFM Heisenberg model. The ground state energy now depends on the distortion as
\begin{equation}
    E(\delta u)_{\mathrm{Heis}}  = E_{\mathrm{GS}} + K (\delta u)^2 - c (g\,\delta u)^{4/3}
\end{equation}
where $c$ is a constant. 
Due to the non-analytic behavior of the spin energy as a function of distortion, 
there is a solution that minimizes the free energy with a finite distortion $\delta u_{\mathrm{eq}} \sim (c\,g^\chi/K)^{1/(2-\chi)}.$
For $\chi < 2$, this state has lower energy than $\delta u =0$. Hence, at zero temperature $T=0$ and in the thermodynamic limit, a lattice distortion and concomitant spin dimerization is induced, as has additionally been confirmed by numerical DMRG simulations \cite{papenbrock03,PhysRevB.75.052404}.

At finite temperatures, the spin-Peierls instability competes with thermal fluctuations. Above a spin-Peierls temperature $T_c = 0.253 \, J[g^2/(\mathsf{a}^2 K)]$, no instability occurs \cite{PhysRevB.70.214436}.
It has widely been measured in experiments in materials that realize such quasi-one dimensional spin chain models~\cite{buzdin1980spin,nishi1994neutron}.

Furthermore, if one considers lattice distortions as quantum mechanical optical phonons, the spin-Peierls instability moves to finite coupling \cite{giamarchi2004quantum}.
Optical phonons $u(x,t)$ with energy $\omega_0$ couple to the bosonized spins in the same way via $\int\dd[2]{x} g\, u(x) \sin(2\phi)$. 
They generate the interaction $-g^2 \int\dd[2]{x} \cos(4\phi)$ and kinetic terms when the high-energy modes are integrated out, shifting the value of $V_4$ and the Luttinger parameter. At sufficient coupling, the induced effective interaction $V_4'$ becomes relevant; a mass gap is induced and the spin-liquid destabilized above a critical coupling $g_c\sim\omega_0^{3/2}$ \cite{citro05}.

\section{U(1) Dirac spin liquids}\label{sec:DSL}

\begin{figure}
\centering
\includegraphics[width=0.9\columnwidth]{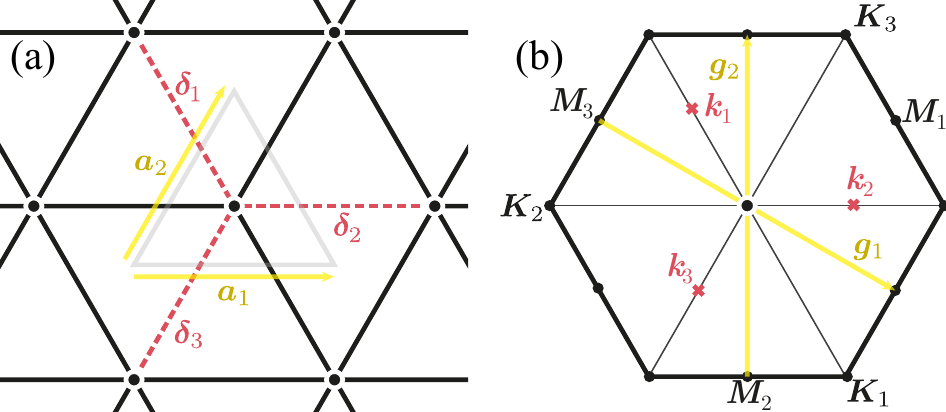}
\caption{(a) Triangular lattice with unit vectors $\bvec{a}_i$ and nearest neighbour bonds $\bvec{\delta}_i$ highlighted. (b) Reciprocal lattice with inverse lattice vectors $\bvec{g}_i$ and high-symmetry points in the Brillouin zone $\bvec{K}_a$ and $\bvec{M}_a$ labeled. The points $\bvec{k}_a = -\bvec{K}_a/2$ are the momenta eigenvalues of gapless monopole excitations \cite{song19}.} 
\label{fig2}
\end{figure}

In the following, for concreteness, we focus on the DSL on a triangular lattice, which may be realized in the $J_1$--$J_2$ Heisenberg model
\begin{equation}\label{eq:heisj1j2}
    H = J_1 \sum_{\left<i,j\right>} \vec{S}_i\cdot \vec{S}_j
    +J_2 \sum_{\left<\left<i,j\right>\right>} \vec{S}_i\cdot \vec{S}_j,
\end{equation}
where $J_1$ and $J_2$ are nearest- and next-nearest-neighbour antiferromagnetic exchange couplings.
Numerical studies indicate that a spin liquid phase is realized around the classically-disordered point of $J_2/J_1 = 1/8$ \cite{PhysRevB.42.4800}, with recent numerical evidence pointing towards a $\Uone$ Dirac spin liquid \cite{PhysRevLett.123.207203, Sherman2022, drescher2022dynamical}.
For reference, we define the triangular lattice vectors 
\begin{equation}
	\bvec a_1 = \mathsf{a} (1,0)^\top \quad \text{and} \quad \bvec a_2 = \mathsf{a}(1/2,\sqrt{3}/2)^\top,
\end{equation}
where $\mathsf{a}$ denotes the lattice constant,
and the nearest-neighbor vectors are labeled $\bvec{\delta}_1 = \bvec{a}_2-\bvec{a}_1$, $\bvec{\delta}_2 = \bvec{a}_1$ and $\bvec{\delta}_3 = -\bvec{a}_2$ as depicted in Fig.~\ref{fig2}.
The corresponding reciprocal lattice vectors are
\begin{equation}
	\bvec g_1 = (2\pi,-2 \pi/\sqrt{3})^\top/\mathsf{a} \quad \text{and} \quad\bvec g_2 = (0, 4\pi/\sqrt{3})^\top/\mathsf{a}.
\end{equation}
The high-symmetry points of the  hexagonal Brillouin zone are indicated in Fig.~\ref{fig2}, and the $\bvec{K}_a$ points are explicitly
\begin{equation}
    \bvec{K}_{1,3}=[4\pi/3\mathsf{a}](1/2,\mp\sqrt{3}/2)
     \quad \text{and} \quad
     \bvec{K}_{2}=[4\pi/3\mathsf{a}](-1,0).
\end{equation}

\subsection{QED$_3$ effective theory}

In a parton mean-field description, one can rewrite spin-operators in terms of fermionic partons $f_{i,\tau}$ as $\vec S_i = f_{i,\alpha}^\dagger \vec{\sigma}_{\alpha \beta} f_{i,\beta}$ and perform a subsequent decoupling of spin-spin interactions in the microscopic Hamiltonian \eqref{eq:heisj1j2}.
The fermionic parton construction introduces unphysical degrees of freedom which can be projected out by enforcing the unit-occupancy constraint $f_{i,\uparrow}^\dagger f_{i,\uparrow} + f_{i,\downarrow}^\dagger f_{i,\downarrow} \overset{!}{=}1$ on each site. This leads to the emergence of a \emph{dynamical} U(1) gauge field $a_{ij}$.
The DSL state is then defined by the mean-field Hamiltonian
\begin{equation}
    \mathcal{H}_\mathrm{DSL} = \sum_{\alpha = \uparrow, \downarrow} \sum_{i,j} \chi_{ij} f_{i,\alpha}^\dagger f_{j,\alpha}
\end{equation}
where the ansatz $\chi_{ij}$ consists of the fermions (spinons) hopping in a staggered $\pi$-flux background of the dynamical gauge field \cite{iqbal16}.
Under $\Uone$ gauge transformations, the fermions transform as $f_j \to \eu^{\iu \theta_j} f_j$, the gauge field transforms as $a_{ij} \to a_{ij} - \theta_i + \theta_j$, and the mean-field Hamiltonian as $\chi_{ij} \mapsto \chi_{ij} \eu^{\iu a_{ij}}$.
Note that this implies that $a_{ij} \in [0,2 \pi]$; the gauge group is \emph{compact}.

The spectrum of the resulting fermion Hamiltonian features two Dirac cones per spin at two distinct momenta in the hexagonal Brillouin zone. 
At low energies (long wavelengths),
this gives rise to an additional $\SUtwo$ valley degree of freedom;
one can pass over to a continuum description, with the Lagrangian (in Euclidean signature)
\begin{equation} \label{eq:qed-3}
  \mathcal{L}_{\mathrm{QED}_3} = -
  \sum_{i=1}^4 \overline{\psi}_i \gamma^\mu (\partial_\mu - \iu a_\mu ) \psi_i
  +\frac{1}{4 g^2} f_{\mu\nu} f^{\mu \nu},
\end{equation}
describing Dirac fermions with $N=4$ flavors (2 spin 
$\times$ 2 valley) minimally coupled to a U(1) gauge field $a_\mu$ with field-strength tensor $f_{\mu \nu}$, thus corresponding to Quantum Electrodynamics in 2+1-dimensions. 

Note that in the low-energy theory $\mathcal{L}_{\mathrm{QED}_3}$, the spin- and valley symmetries $\SUtwo_\mathrm{spin} \times \SUtwo_\mathrm{valley}$ become enhanced to a global $\SUfour$ symmetry acting on the flavor degrees of freedom of the Dirac fermions.
Further, the theory possesses Lorenz symmetry as well as discrete charge conjugation $\mathcal{C}$, time reversal $\mathcal{T}$ and parity $\mathcal{P}$ symmetries.

In 2+1 dimensions, the coupling in Eq.~\eqref{eq:qed-3} has mass dimension $[g^2]=1$, implying that the theory is strongly coupled in the infrared (IR), long-wavelength, limit.
For a sufficiently large number of flavors $N$, \QED3 is believed to flow to a strongly-coupled IR-fixed point with conformal symmetry \cite{appel86,nash89}.
In the remainder of this work we assume that this fixed point is realized also for $N=4$ \cite{bashir08,herbut16} and that spontaneous chiral symmetry breaking does not occur \cite{apple88,grover14,janssen14} for Eq.~\eqref{eq:heisj1j2}.

\subsection{Monopole operators}

As written above, the low-energy theory $\mathcal{L}_{\mathrm{QED}_3}$ appears to have an additional global symmetry $\Uone_\mathrm{topo}$ with the conserved current $j_{\mathrm{topo}}^\mu = \frac{1}{2 \pi} \epsilon^{\mu \nu \rho} \partial_\nu a_\rho$, corresponding to the conservation of magnetic flux (hence sometimes referred to as `magnetic symmetry').
However, given that $\mathcal{L}_{\mathrm{QED}_3}$ emerges as the low-energy description of a \emph{compact} $\Uone$ lattice gauge theory, instanton events may add/remove integer-$q$ multiples of $2 \pi$ magnetic flux.
The corresponding operators are referred to as monopole operators $\mathcal{M}^\dagger_{2 \pi q}$ \cite{boro02}, and are charged under the $\Uone_\mathrm{topo}$.
These monopole operators do not possess an explicit representation as some polynomial in the gauge-field $a_\mu$ and spinons $\psi$ \cite{boro02,song19}, but rather correspond to topologically non-trivial gauge field configurations \cite{chester16}.
From the lattice perspective, this $2\pi$ flux insertion can be, for example on the torus, understood as modifying the $\pi$-flux mean-field state by spreading a uniform $2\pi/L^2$ flux in each unit cell \cite{alex2023quantum,budaraju2023piercing}. 

It is well known that pure compact $\Uone$ gauge theory is unstable towards confinement due to the proliferation of monopoles \cite{polyako77}.
In the context of the U(1) gauge theory coupled to $N=4$ gapless fermions, the question of stability is more involved.
The monopoles can be schematically written as $\Phi^\dagger_{a} \sim f^\dagger f^\dagger \mathcal{M}_{2 \pi}^\dagger$ with $a=1\dots6$ corresponding to six ways of half-filling four Dirac zero modes $f$, as mandated by gauge invariance \cite{song19}.
These monopoles transform under a six-dimensional representation of $\SUfour$.
Understanding the stability of the system requires understanding if monopole excitations $\Phi_a$ are (1) relevant perturbations to the IR fixed point \emph{and} (2) allowed by the microscopic symmetries of the system:
\begin{enumerate}[label=(\arabic*)]
\item Both analytical \cite{chester16} as well as numerical studies \cite{karthik16} point towards monopoles being strongly relevant for \QED3 with $N=4$ flavors, with a recent bootstrap \cite{alba22} calculation suggesting a monopole scaling dimension $\Delta_\Phi \simeq 1.02$ (we adopt the convention that $[\Phi] \sim L^{-\Delta_\Phi}$ with some length scale $L$, e.g. the system size, so that $\Delta_\Phi < 2+1$ implies relevance in 2+1 dim.).

\item On the triangular lattice all six monopole operators have non-trivial quantum numbers under the UV symmetries of the system \cite{song19,song20}. Hence, as long as these lattice symmetries are preserved, single-monopoles are not allowed and no ordering instability via monopole proliferation will occur.
In contrast, on the square and honeycomb lattice, there exists a trivial (symmetry-allowed) monopole operator, such that the DSL state on these lattices is unstable.
\end{enumerate}

We include a brief overview of the application of conformal field theory to the description of the DSL IR fixed point for completeness in Appendix~\ref{sec:CFT-intro}. The power of this approach comes from being able to describe the fixed point by a \emph{spectrum of primary operators} $\mathcal{O}_i$, with the knowlege of their scaling dimensions $\Delta_i$ and symmetry transformation properties. By conformal invariance, two-point correlation functions (in Euclidean spacetime) of primaries can then be written exactly as
\begin{equation} \label{eq:cft-correl}
    \langle \mathcal{O}_i(x) \mathcal{O}_j(y) \rangle = \frac{\delta_{ij}}{|x-y|^{2\Delta_i}}.
\end{equation}
The conformal theory is fully described by this spectrum of operators alongside an understanding of how they combine --- this is provided by the operator product expansion (OPE).
Within our scheme we focus on the most-relevant charge-one monopoles $\Phi$ and the leading-order OPE can be summarized as $\Phi^\dagger\times\Phi \sim \mathds{1}+\cdots$.

While the above discussion was focused on monopole proliferation, there may exist other perturbations that constitute relevant deformations to the IR fixed point, such as fermion bilinears (see also Appendix~\ref{sec:CFT-intro}). 
However, as discussed in Refs.~\cite{song19,song20}, on the triangular lattice, these perturbations are also either irrelevant or symmetry-forbidden. 
The leading symmetry-allowed operator is a monopole which inserts $6\pi$ topological flux and is expected to be irrelevant; as such the DSL on the triangular lattice is expected to be stable.

\section{Monopole-lattice coupling}\label{sec:coupling}

In this section, we derive the effective theory of a lattice distortions coupled to monopole operators in \QED3. This is based on a symmetry analysis facilitated by the knowledge of the symmetry quantum numbers of monopole operators on the triangular lattice \cite{song19,song20,galitski23}.

\subsection{Monopole symmetry quantum numbers}

{\setlength{\tabcolsep}{10pt}\renewcommand{\arraystretch}{1.2}\begin{table}
\caption{\label{tab:UV-sym-mono}Discrete symmetry transformations of monopole operators.}
\begin{tabular}{lccccc}
 \hline \hline
     & $T_j$ & $R$ & $C_6$ & $\mathcal{T}$ \\ \hline
 \noalign{\vskip 0.5mm}
     $\Phi_1$ & $\eu^{-\iu \bvec{k}_1\cdot \bvec{a}_j} \Phi_1 $ & $-\Phi_3$ & $\Phi_2^\dagger$ & $\Phi_1^\dagger$ \\ 
     $\Phi_2$ & $\eu^{-\iu \bvec{k}_2\cdot \bvec{a}_j} \Phi_2 $ & $\Phi_2$ & $-\Phi_3^\dagger$ & $\Phi_2^\dagger$ \\ 
     $\Phi_3$ & $\eu^{-\iu \bvec{k}_3\cdot \bvec{a}_j} \Phi_3 $ & $-\Phi_1$ & $-\Phi_1^\dagger$ & $\Phi_3^\dagger$ \\
 \noalign{\vskip 1mm}
     $\Phi_{a=4,5,6}$ & $\eu^{-\iu \bvec{K}_{a-3}\cdot \bvec{a}_j}  \Phi_{a}$  & $\Phi_{a}$ & $-\Phi_{a}^\dagger$ & $-\Phi_{a}^\dagger$\\
\noalign{\vskip 0.5mm}
      \hline \hline
\end{tabular}
\end{table}}

The monopoles $\Phi_{a}$ form a six-dimensional representation of the IR-symmetry group $\SUfour$.
The two (independent) microscopic symmetries $\SUtwo_\mathrm{valley}$ and $\SUtwo_\mathrm{spin}$ as subgroups of $\SUfour$ correspond to mutually commuting $\SOthree_\mathrm{valley}$ and $\SOthree_\mathrm{spin}$ subgroups of $\SO(6)$, such that we can decompose $\Phi_a$ into `valley-triplet, spin-singlet' ($a=1,2,3$) and `valley-singlet, spin-triplet' components ($a=4,5,6$).

Importantly, microscopic (lattice) UV symmetries such as translations, discrete rotations, and reflections are embedded in the enlarged symmetry group of the IR theory (in other words, microscopic symmetries are not broken in the flow to the IR fixed point).
The monopole operators $\Phi_a$ carry non-trivial quantum numbers under lattice symmetries, which have been determined in Refs.~\cite{song19,song20}.
In Tab.~\ref{tab:UV-sym-mono} we reproduce the transformation of monopoles on the triangular lattice under discrete lattice translations $T_j$ by unit vectors $\bvec{a}_j$, reflections $R$ in the vertical direction, discrete rotations $C_6$, and time reversal $\mathcal{T}$.
The three valley-triplet monopoles ($a=1,2,3$) carry the same quantum numbers as order parameters of valence-bond solids on the triangular lattice with lattice momenta $\bvec{k}_a = -\bvec{K}_a/2$, respectively. Hence, we can write (up to a global multiplicative constant not fixed by symmetry)
\begin{equation}\label{eq:vbsmonopolemapping}
    \vec{S}_i \cdot \vec{S}_{i+\bvec{\delta}_a} \simeq s_a \Re[ \Phi_{a}(\bvec{r}_i) \eu^{\iu \bvec{k}_a \cdot \bvec{r}_i}],
\end{equation}
where we introduce the sign factors $s_a = (+1,+1,-1)$ which are needed to reproduce the relative negative sign for $\Phi_3$ in Tab.~\ref{tab:UV-sym-mono}.
The three spin-triplet monopoles ($a=4,5,6$) transform identically as order parameters for antiferromagnetic $120^\circ$ Néel order which determine the spin density as
\begin{equation} \label{eq:spin-monopole}
    S_i^\alpha  \simeq \Re[\iu \Phi_{3+\alpha}(\bvec{r}_i) \eu^{\iu \bvec{K}_\alpha \cdot \bvec{r}_i}],
\end{equation}
where $\alpha=x,y,z$ (1,2,3) denotes the three $\SUtwo_\mathrm{spin}$ components.
(Note that the LHS in Eqs.~\eqref{eq:vbsmonopolemapping} and \eqref{eq:spin-monopole} only correspond to the ``lowest-order'' microscopic expressions that transform as the RHS under symmetry transformations, and one may find terms of multiple spin operators that exhibit the same symmetry operators.)
These expressions gives way to the interpretation of the monopoles as \emph{disorder operators} \cite{boro02}, with their proliferation $\langle\Phi^a \rangle \neq 0$ yielding conventionally ordered phases.
Crucially, the nature of the phase is determined by the proliferating components of $\Phi_a$ and the corresponding broken global symmetries.

\subsection{Lattice deformation}

We first neglect any quantum dynamics of lattice distortions and focus on classical displacement fields.
Considering actual materials, this approximation is justified if phonons have a large inertial mass $m_0 \to \infty$ such that the phonon frequency vanishes $\omega_0 \to 0$.
In the literature on the spin-Peierls transition in one-dimensional systems, this is commonly referred to as the static phonon limit or the `adiabatic approximation' \cite{cross79,orignac04,citro05}.

Our starting point is the observation that the valley-triplet (spin-singlet) monopoles $\Phi_a$ ($a=1,2,3$), which act as order parameters for VBSs, have lattice momenta $\bvec{k}_a = -\bvec{K}_a/2$.
Using Tab.~\ref{tab:UV-sym-mono}, we note that the following deformation to the DSL fixed point is symmetry-allowed,
\begin{equation} \label{eq:h-phi-r}
    \mathcal{H} \sim \left( \eu^{\iu \bvec{k}_1 \cdot \bvec{R}} \Phi_1 + \eu^{\iu \bvec{k}_2 \cdot \bvec{R}} \Phi_2 - \eu^{\iu \bvec{k}_3 \cdot \bvec{R}} \Phi_3 + \hc \right).
\end{equation}
We emphasize that care must be taken in separating length scales: $\bvec{R}$ ($\bvec{k}_a$) are coordinates (wavevectors) on the order of microscopic length scales such as the lattice constant $\mathsf{a}$, where the $\Phi_a$ transform as given in Tab.~\ref{tab:UV-sym-mono}.
The scaling nature of the $\Phi_a$ as primaries in a CFT, and the corresponding power-law form of correlation functions \eqref{eq:cft-correl}  only holds on much longer lengthscales (and, equivalently, sufficiently small momenta), where a low-energy continuum formulation becomes justified.
While Eq.~\eqref{eq:h-phi-r} is symmetry allowed, the oscillating prefactors average out on sufficiently long length scales, such that the perturbation is strongly irrelevant.

We now consider an in-plane \emph{deformation} of the real-space lattice $\bvec{R} \to \bvec{x} = \bvec{R} + \bvec{u}(\bvec{x})$, where $\bvec{u}(\bvec{x})$ is a \emph{displacement field}, as shown in the left panel of Fig.~\ref{fig1}.
Note that we work in implicitly-defined (Eulerian) coordinates of the deformed system $\bvec{x}$.

Expanding Eq.~\eqref{eq:h-phi-r} to the first non-trivial order in $|\bvec{u}(\bvec{x})|/\mathsf{a} \ll 1$ (assuming that lattice distortions are small compared to the lattice constant $\mathsf{a}$), we obtain the monopole-lattice coupling Hamiltonian
\begin{equation}
    \mathcal{H}_g[\bvec{u}(\bvec{x})] = g \sum_{a=1,2,3} s_a \left[ \left( \iu \bvec{k}_a \cdot \bvec{u}(\bvec{x}) \right) \eu^{\iu \bvec{k}_a \cdot \bvec{x}} \Phi_a + \hc \right],
\end{equation}
where we use again the sign factors $s_a = (+1,+1,-1)$ for convenience of notation. We have dropped the 0-th order terms (with oscillating phases) in the expansion.
Note that $\iu \bvec{k}_a\cdot \bvec{u}$ transforms as a scalar under point-group operations $R$ and $C_6$ and is appropriately even under time reversal $\mathcal{T}$; therefore the prefactor retains the full symmetry of the undistorted system Eq.~\eqref{eq:h-phi-r}.
With Eq.~\eqref{eq:vbsmonopolemapping}, this interaction can be seen as a modulation of the nearest-neighbor $J_1$ coupling through the difference of site-spacing in the following way:
\begin{equation}\label{eq:bdhopping}
    H_g [\bvec{u}_i] \simeq \tilde{g} \sum_i\sum_{a=1,2,3} \bvec{k}_a \cdot [\bvec{u}(\bvec{r}_i)-\bvec{u}(\bvec{r}_i+\bvec{\delta}_a)] \, \vec{S}_{\bvec{r}_i}\cdot \vec{S}_{\bvec{r}_i+\bvec{\delta}_a}.
\end{equation} 
We can expand the real-component distortion field $\bvec{u}(\bvec{x})$ in eigenstates of the lattice momentum $\bvec{u}_\bvec{Q}$
\begin{equation} \label{eq:fourier-u-x}
    \bvec{u}(\bvec{x}) = \sum_\bvec{Q} \eu^{\iu \bvec{Q} \cdot \bvec{x}} \bvec{u}_\bvec{Q},
\end{equation}
where reality of $\bvec{u}(\bvec{x})$ implies $\bvec{u}_\bvec{Q} = \bvec{u}_{-\bvec{Q}}^\ast$. 
In momentum space, the coupling to monopoles is of the form
\begin{equation} \label{eq:h-g-general}
    \mathcal{H}_g [\bvec{u}(\bvec{x})] = g \sum_{a=1,2,3} \sum_\bvec{Q} s_a \left[ \left( \iu \bvec{k}_a \cdot \bvec{u}_\bvec{Q} \right) \eu^{\iu (\bvec{k}_a +\bvec{Q}) \cdot \bvec{x}} \Phi_a + \hc \right].
\end{equation}

Crucially, only terms with $\bvec{Q} = - \bvec{k}_a$ do not contain any oscillating prefactors and will thus be the ones relevant at the lowest energies; these are
\begin{equation} \label{eq:h-g-explicit}
    \mathcal{H}_g [\bvec{u}(\bvec{x})] = g \sum_{a=1,2,3}s_a \left[ \left( \iu \bvec{k}_a \cdot \bvec{u}^*_{\bvec{k}_a} \right) \Phi_a + \hc \right].
\end{equation}
Our key observation is that, in analogy to the one-dimensional spin-Peierls transition, these terms may precipitate an instability.
Indeed we will go on to use energetic arguments to establish the same result in the two-dimensional DSL.
The response of \QED3 to this perturbation will be dominated by lattice distortions with crystal momentum $\bvec{k}_a$.
We highlight the translational invariance of \eqref{eq:h-g-explicit} by noting that under translation, $\bvec{u}^*_{\bvec{k}_a}\to \eu^{ \iu \bvec{k}_a \cdot \bvec{a}} \bvec{u}^*_{\bvec{k}_a}$, and the coupling is unchanged, just like for the Heisenberg chain (see Appendix~\ref{app:bozonization}).

\subsection{Elastic energy}
We model lattice distortions as having a potential energy that is quadratic in the relative displacement between nearest neighbor sites,
\begin{equation}
    H_\mathrm{ph} = \frac{K}{2} \sum_{\langle ij \rangle} \left| \bvec{u}(\bvec{x}_i) - \bvec{u}(\bvec{x}_j) \right|^2.
\end{equation}
Here, $K$ takes the role of an effective spring constant for the displacement between atoms on the triangular lattice.
Taking the continuum limit, inserting \eqref{eq:fourier-u-x},  
and decomposing into transverse and longitudinal polarizations $\bvec{u}_{\bvec{q}}=\sum_{s=t,l}\bvec{\epsilon}_{\bvec{q},s}u_{\bvec{q},s}$ yields the Hamiltonian
\emph{density}
\begin{multline}\label{eq:distcost}
    \mathcal{H}_\mathrm{ph}[\bvec{u}] = \sum_{\bvec{q}} \mathcal{K}_{\bvec{q}}  | \bvec{u}_{\bvec{q}} |^2
    = \sum_{\bvec{q},s}\mathcal{K}_{\bvec{q}}  | u_{\bvec{q},s} |^2
    \\
    \quad
    \mathcal{K}_{\bvec{q}} = \frac{2K}{3\mathsf{a}^2}\left(3-\left[\cos ( \mathsf{a} q_x) + 2 \cos(\frac{\mathsf{a} q_x}{2}) \cos(\frac{\sqrt{3} \mathsf{a} q_y}{2})\right]\right).
\end{multline}
The polarization vectors $\bvec{\epsilon}_{\bvec{q},s}$ are orthonormal, where the longitudinal direction is defined as $\bvec{\epsilon}_{\bvec{q},l} = \bvec{q}/|\bvec{q}|$.
Importantly, the complex scalar modes $u_{\bvec{q},s}$ are independent and degenerate, meaning the Hamiltonian of the longitudinal modes can be seperated.
The energy density cost of a distortion goes to zero at the center of the Brillouin zone and is of order $K/\mathsf{a}^2$ towards the zone edges.

\section{DSL spin-Peierls instability}\label{sec:instability}
\subsection{Conformal perturbation theory} \label{sec:conf-pert-theory}

\subsubsection{IR Regularization}

Our main goal is to study whether the phonon-monopole coupling generates an instability. To this end, we compute the energy of \QED3 in the background of arbitrary displacement fields $\bvec{u}(x)$ and extremize the resulting energy functional.
It is convenient to employ a path-integral formulation, where we obtain a functional for the effective energy density $\mathcal{E} = E/V$ (per volume $V$) via the zero-temperature limit of the free energy,
\begin{equation}
    \mathcal{E}_{\mathrm{QED}_3}[\bvec{u}] = \frac{1}{V} \lim_{\beta \to \infty} F_{\mathrm{QED}_3}[\bvec{u}].
\end{equation}
The free energy of \QED3 coupled to some background displacement field $\bvec{u}(x)$,
\begin{equation}
    F_{\mathrm{QED}_3}[\bvec{u}] = 
    -\frac{1}{\beta} \log \mathcal{Z}_{\mathrm{QED}_3}^{S^1_\beta \times \Sigma}[\bvec{u}],
\end{equation}
is given in terms of the partition function
\begin{equation} \label{eq:z-qed-u}
    \mathcal{Z}_{\mathrm{QED}_3}^{S^1_\beta \times \Sigma}[\bvec{u}] = \int \mathcal{D}[\{\mathcal{O}_\mathrm{CFT} \}] \, \eu^{- S_{\mathrm{QED}_3} - S_g[\bvec{u}] }
\end{equation}
with the respective actions $S_A= \int_0^\beta \du \tau L_A$ on the manifold $S^1_\beta \times \Sigma$.
Here, $S^1_\beta$ corresponds to a circle in the imaginary time direction with circumference inverse temperature $\beta = 1/T$, and $\Sigma$ is some spatial manifold (for example a 2-sphere with radius $L$, $S^2_L$). We mostly focus on the thermodynamic limit $\Sigma = \lim_{L\to \infty} S^2_L = \mathbb{R}^2$ (note that the partition function $\mathcal{Z} = \mathcal{Z}^{S^3_L}$ may also be defined on a 3-sphere with radius $L$, see, e.g.,~Ref.~\onlinecite{luo22}, making the conformal $\SO(3)$ symmetry group manifest).

Away from the large-$N$ limit, \QED3 is not solvable, and thus $\mathcal{Z}_{\mathrm{QED}_3}$ cannot be computed exactly.
Instead, we use \eqref{eq:z-qed-u} mainly as a computational framework.
The schematic notation of the measure of the path integral indicates that all operators at the \QED3 CFT are to be integrated over. We will not attempt a rigorous definition here, as there is no explicit form for $S_{\mathrm{QED}_3}$ in terms of CFT operators available. 

\subsubsection{Weak coupling analysis}
In the limit of weak coupling we can work perturbatively and exploit the fact that two-point functions at the CFT fixed point are known. We assume that the monopole-lattice action is a small perturbation, justified by assuming small coupling $g$ and by our previous assumption that $|\bvec{k}_a\cdot \bvec{u}_{\bvec{Q}}|\ll1$.
Expanding the Boltzmann weight to quadratic order and taking the logarithm, we have
\begin{multline}
    \log \mathcal{Z}_{\mathrm{QED}_3}[\bvec{u}] = \log \mathcal{Z}_{\mathrm{QED}_3} - \langle S_g \rangle_{\mathrm{QED}_3} \\
    + \frac{1}{2} \left( \langle S_g^2 \rangle_{\mathrm{QED}_3} - \langle S_g \rangle_{\mathrm{QED}_3}^2  \right) + \dots,
\end{multline}
where the expectation values are to be taken with respect to the path integral \eqref{eq:z-qed-u} with $\bvec{u} \equiv 0$.
At finite temperatures $\beta < \infty$, one-point functions of conformal primaries are generically non-zero $\langle \mathcal{O} \rangle \sim \beta^{-\Delta_\mathcal{O}}$ \cite{iliesu18}, but, importantly, vanish in the zero-temperature limit $\beta \to \infty$, as also mandated by conformal invariance on $\mathbb{R}^3$.
Because we will primarily focus on the zero-temperature limit, we henceforth take $\langle S_g \rangle_{\mathrm{QED}_3} = 0$.

The first non-trivial contribution to $\mathcal{E}_{\mathrm{QED}_3}[\bvec{u}]$ thus occurs at quadratic order (we use the notation $x = (\tau_x,\bvec{x})$ for vectors in 2+1-dim. Euclidean spacetime),
\begin{widetext}
\begin{align}
    \langle S_g^2 \rangle_{\mathrm{QED}_3} &= g^2 \sum_{a,b} s_a s_b \sum_{\bvec{Q},\bvec{Q}'} \int_{S^1_\beta \times \Sigma} \du^3 x  \du^3 y \left[ (\iu \bvec{k}_a \cdot \bvec{u}_{\bvec{Q}})  \eu^{\iu (\bvec{k}_a + \bvec{Q}) \cdot \bvec{x}} (- \iu \bvec{k}_b \cdot \bvec{u}_{\bvec{Q}'}^\ast) \eu^{-\iu (\bvec{k}_b + \bvec{Q}') \cdot \bvec{y}} \langle \Phi_a(x) \Phi_b^\dagger(y) \rangle_{\mathrm{QED}_3} + \hc \right] \nonumber\\
    &= g^2 \beta V \sum_{a} \int_0^\beta \du \tau_{x'} \int_\Sigma \du^2 \bvec{x}' \left[ |\bvec{k}_a \cdot \bvec{u}_\bvec{Q}|^2 \eu^{\iu (\bvec{k}_a + \bvec{Q}) \cdot \bvec{x}'} \langle \Phi_a^\dagger(x') \Phi_a(0) \rangle_{\mathrm{QED}_3} + \hc \right], \label{eq:before-ope}
\end{align}
\end{widetext}
where $V = \mathrm{vol}(\Sigma) = 2 \pi L^2$ is a trivial factor of volume, mandated by the extensiveness. The thermodynamic limit $\Sigma = \mathbb{R}^2$ is obtained by taking the linear dimension $L\to \infty$.
We have used in the above that monopole-monopole two-point functions $\langle \Phi_a(x) \Phi_b(x) \rangle_{\mathrm{QED}_3} = 0$ vanish, monopole-antimonopole correlation functions are diagonal in the monopole flavor index, changed coordinates $x' = x+ y$ and assumed translation invariance $\langle \Phi^\dagger(x+y) \Phi(y) \rangle_{\mathrm{QED}_3} = \langle \Phi(x)^\dagger \Phi(0) \rangle_{\mathrm{QED}_3}$.

For $\bvec{Q} \neq - \bvec{k}_a$ there exist rapidly oscillating terms in Eq.~\eqref{eq:before-ope} (with momentum $\bvec{k}_a + \bvec{Q}$), which, assuming that the monopole two-point function varies sufficiently slowly, produce \emph{finite} contributions to $\langle S_g^2 \rangle_{\mathrm{QED}_3}$ which average out at sufficiently long length scales (in the continuum limit).

The term with $\bvec{Q} = - \bvec{k}_a$ in Eq.~\eqref{eq:before-ope} hence is the dominant contribution at low energies, which is due to the (relevant) coupling between monopoles and displacements at lattice momentum $\bvec{k}_a$, as was written before in Eq.~\eqref{eq:h-g-explicit}.
This justifies us taking this interaction as the starting point for all further analysis.

We caution the reader that, while it may be tempting to immediately use the explicit expression for the conformal two-point function in Eq.~\eqref{eq:before-ope}, this is only justified at lowest energies, when there is a separation between microscopic (lattice) scales and slowly varying continuum fields, which is warranted in Eq.~\eqref{eq:before-ope} \emph{only} for $\bvec{Q}=-\bvec{k}_a$ (otherwise, there is a finite transfer of lattice momentum).

The explicit form of the conformal two-point monopole correlation function in flat space reads $\langle  \Phi_a^\dagger(x) \Phi_b(0) \rangle_{\mathrm{QED}_3} = \delta_{a,b} |x|^{-2\Delta_\Phi}$ [see also Eq.~\eqref{eq:cft-correl}].
As noted in Ref.~\cite{iliesu18}, the cylindrical geometry is conformally flat in the $L\to \infty$ limit, and OPEs converge for $\sqrt{\tau^2+ |\bvec{x}^2|} < \beta \to \infty$, such that we may use above form of the 2-pt. function to do the remaining integral. This integral is divergent in the thermodynamic, zero temperature limit, $\beta, L \to \infty$.
At any other wavevector $\bvec{Q} \neq -\bvec{k}_a$, the exponential in \eqref{eq:before-ope} cuts off the divergent integral and no instability occurs.

We regulate this divergence by working at finite temperatures $\beta < \infty$, which allows us to take the thermodynamic limit $L\to \infty$,
\begin{align}
    \frac{\langle S^2_g \rangle_{\mathrm{QED}_3}}{\beta V} &= 2 g^2 \sum_a \int_0^\beta \du \tau_x \int_\Sigma \du^2 \bvec{x} \,  \frac{| \bvec{k}_a \cdot \bvec{u}_{\bvec{k}_a}|^2 }{|\tau_x^2 + |\bvec{x}|^2|^{\Delta_\Phi} } \nonumber\\
    &= c_{\Delta_\Phi} g^2 \beta^{3-2\Delta_\Phi}  \sum_a | \bvec{k}_a \cdot \bvec{u}_{\bvec{k}_a}|^2 \label{eq:sg2}
\end{align}
where we have introduced the numerical constant $c_{\Delta_\Phi} = 2\pi/ [(\Delta_\Phi-1)(3-2\Delta_\Phi)] > 0$ for $1 < \Delta_\Phi < 3/2$. For current estimates of the monopole scaling dimension this is true, and indeed this leads to a large numerical value $c_{\Delta_\Phi} \gg 1$ because of the proximity of the monopole scaling dimension to one $(\Delta_\Phi-1)\ll 1$ \cite{chester16,karthik16,alba22}.

\subsubsection{Spin-Peierls instability} \label{sec:sp-instab}

We can now obtain the effective energy as a sum of the intrinsic potential energy cost of deformations at wavevectors $\bvec{k}_a$ and the relative energy gain of the DSL by coupling to the displacement field,
\begin{multline}
    \mathcal{H}_{\mathrm{eff}}[\bvec{u}] = \mathcal{H}_\mathrm{ph}[\bvec{u}] + 
    \left(\mathcal{E}_{\mathrm{QED}_3}[\bvec{u}] - \mathcal{E}_{\mathrm{QED}_3}[0] \right) \\=  \sum_{a=1,2,3} \left( \kappa |\bvec{u}_{\bvec{k}_a}|^2 - \lim_{\beta \to \infty}  c_{\Delta_\Phi} g^2 \beta^{3-2 \Delta_\Phi}  | \bvec{k}_a \cdot \bvec{u}_{\bvec{k}_a}|^2  \right) \label{eq:perturbative-energy}
 \end{multline}
where we introduce $\kappa = \mathcal{K}_{\bvec{k}_a}$.
We observe that the second term in the parenthesis is minimized by taking lattice displacements that are \emph{longitudinal} along the momenta $\bvec{u}_{\bvec{k}_a} = u_a\, \bvec{k}_a / | \bvec{k}_a|.$
The scalar amplitudes $u_a$ are the longitudinal components $u_{\bvec{k}_a,l}$ at the three independent displacement vectors $\bvec{k}_a$.
Henceforth we use the vector notation for these longitudinal displacements $\vec{u}=(u_1,u_2,u_3)^\top$, such that we can express the effective potential in the form
\begin{equation}\label{eq:short-energy-u}
    \mathcal{H}_{\mathrm{eff}}[\vec{u}] = \left(\kappa - c_{\Delta_\Phi} g^2 |\bvec{k}_a|^2 \beta^{3-2\Delta_\Phi}\right)|\vec{u}|^2.
\end{equation}

In the zero-temperature limit $T = \beta^{-1} \to 0$ the effective energy density is not bounded from below for any finite (small) coupling $g>0$, and hence the energy density diverges $\mathcal{H}_{\mathrm{eff}}\to -\infty$ for any $u_a\neq 0$. It is thus energetically preferable for the system to acquire a lattice distortion.
We thus conclude that the coupling of valley-triplet monopoles (acting as VBS order parameters) to longitudinal displacement modes with wavevectors $\bvec{k}_a$ induces a weak-coupling instability of the DSL at low temperatures, analogous to the 1D spin-Peierls mechanism discussed in Section~\ref{sec:luttsp}.

The critical temperature scale $T_{\mathrm{SP}} = 1/\beta_c$ for the ordering instability is obtained by analyzing where the quadratic potential $\mathcal{H}_{\mathrm{eff}}[\vec{u}]$ [Eq.~\eqref{eq:short-energy-u}] changes sign, yielding $\kappa = c_{\Delta_\Phi} g^2 \beta_c^{3-2\Delta_\Phi} |\bvec{k}_a|^2$. Recall that both the stiffness $\kappa$ and the lattice momentum $\bvec{k}_a$ scale as the inverse of the lattice constant;
we hence evaluate the critical spin Peierls temperature to be
\begin{equation} \label{eq:tsp}
    T_{\mathrm{SP}} \sim \left[ \frac{1}{(\Delta_\Phi-1)} \frac{g^2}{K} \right]^{1/(3-2\Delta_\Phi)}.
\end{equation}
The VBS ground state is therefore separated from a high-temperature DSL phase by a finite-temperature spin-Peierls transition, again in analogy to the 1D case.

\subsection{Scaling ansatz} \label{sec:scaling}

In the above calculation, the weak-coupling approximation and expansion of the partition function apply when $g^2 |\bvec{k}_a|^2 \beta^{3-2\Delta_\Phi} |\vec{u}| \ll 1$.
For stronger coupling $g$ or larger $\beta\to\infty$ (or equivalently $L\to\infty$ in the symmetric regularization), the first order perturbative expansion is not sufficient.
In the following section, we take a complementary approach to the phonon-induced ordering instability by making a scaling approximation for the energy gain due to deforming QED$_3$ by the coupling of monopoles to the classical displacement field. This applies in the opposite limit of strong coupling $g^2 |\bvec{k}_a|^2\beta^{3-2\Delta_\Phi} |\vec{u}| \gg 1$.

We write the deformation to the \QED3 action in the symmetric form $\mathcal{S}_g[\vec{h}]= g \sum_a h_a^*\Phi_a + \hc$ with non-zero $h_a = \iu s_a |\bvec{k}_a|\,u_a$ for $a=1,2,3$.
From power-counting at the fixed point, we find a strong-coupling action that is scale-invariant and compatible with the symmetries at the fixed point as
\begin{equation} \label{eq:h-scaling}
    \mathcal{S}_\mathrm{sc}[\vec{h}] = - \left(\frac{2 c_\mathrm{sc}}{\chi}\right) |g\,\vec{h}|^\chi
\end{equation}
with $\chi = 3 / (3-\Delta_\Phi) > 0$ and $c_\mathrm{sc} > 0$ an undetermined positive scaling constant \cite{luo22}.
This implies that in the scaling regime, we can write the deformation-induced change in the energy density as
\begin{equation} \label{eq:delta-e-scaling}
    \Delta\mathcal{E}_{\mathrm{QED}_3}[\bvec{u}] \equiv \mathcal{E}_{\mathrm{QED}_3}[\bvec{u}] - \mathcal{E}_{\mathrm{QED}_3} 
    \approx - \frac{2 c_\mathrm{sc}}{\chi} \big| g |\bvec{k}_a| \big|^\chi | \vec{u} |^\chi.
\end{equation}
This non-analytic contribution to the ground-state energy must outcompete the (leading) quadratic energy cost for the longitudinal displacements, $\mathcal{H}_\mathrm{ph} = \kappa |\vec{u}|^2$. 
Comparing with Eq.~\eqref{eq:delta-e-scaling}, we find that it is energetically preferable for the system to undergo a lattice distortion with amplitude $|\vec{u}| \neq 0$ as long as $0 < \chi < 2$.
This implies that the instability occurs for monopole scaling dimensions $\Delta_\Phi < 3/2$, which is satisfied by the assumed value $\Delta_\Phi \approx 1.02$ \cite{alba22}.

Within this scaling ansatz, we can further estimate the magnitude of the induced displacement field by minimizing $\mathcal{H}_\mathrm{ph} + \Delta\mathcal{E}_{\mathrm{QED}_3}[\bvec{u}]$ with respect to $|\vec{u}|$, and obtain
\begin{equation}
    |\vec{u}| 
    = \left| \frac{2 c_\mathrm{sc}}{\kappa} \right|^\frac{3-\Delta_\Phi}{3-2\Delta_\Phi} \left(g |\bvec{k}_a| \right)^\frac{3}{3-2\Delta_\Phi}.
\end{equation}

\subsection{Symmetry of the distortion field}\label{sec:symdistfield}

Both the perturbative approach in Sec.~\ref{sec:conf-pert-theory} as well the scaling ansatz in Sec.~\ref{sec:scaling} suggest a weak-coupling instability of the Dirac spin liquid coupled to \emph{classical} displacement fields.
Importantly, the respective effective actions in Eqs.~\eqref{eq:perturbative-energy} and \eqref{eq:delta-e-scaling} were given as functions of the norm $|\vec{u}|$ of the complex vector $\vec{u} = (u_1,u_2,u_3)^\top$, seemingly implying invariance under global $G\in \mathrm{U}(3)$ transformations $\hat{u} \mapsto G \hat{u}$.
However, this global continuous symmetry is \emph{accidental} as it does not correspond to the physical symmetry group of microscopic lattice symmetries (consisting of translations, sixfold rotation $C_6$ and mirror $R$).

This accidental degeneracy has two distinct origins and thus will be lifted by two separate mechanisms. In the following, we argue that the resultant lattice distortion is $C_3$ symmetric, and use this result to derive the microscopic form of the lattice distortion.

\subsubsection{Lifting of accidental $\mathrm{U}(3)$ symmetry}

We first note that aforementioned $\mathrm{U}(3)$ symmetry acts on the VBS monopoles, which transform as $\SO(3)_\mathrm{valley}$-vectors under the emergent symmetry group of the \QED3 fixed-point action.
We therefore conclude that the global $\mathrm{U}(3)$ redundancy is an artifact of the approximations made to the effective action, in both the perturbative regime as well as using the scaling ansatz. To this end, we note that $2\bvec{k}_1 - 2\bvec{k}_2 = 2\bvec{k}_2 - 2\bvec{k}_3 = 2\bvec{k}_1 - 2\bvec{k}_3 = 0$ up to reciprocal lattice vectors, and hence $| \vec{h} \cdot \vec{h}| $ transforms trivially under the $\SO(6) \times \Uone_\mathrm{top}$ symmetry group.
This term will generically appear in an effective action obtained after `integrating out' the Dirac spin liquid and breaks the $\mathrm{U}(3)$ redundancy down to $\SOthree_\mathrm{valley}$. Importantly, we assume that these terms are small, so that they only act to select some states out of the accidentally-degenerate manifold, but not change the mangitude of the order parameter itself.

To be explicit, one may add a term $\sim |\vec{h}\cdot \vec{h}|^{\chi/4}$ to the scaling form in Eq.~\eqref{eq:h-scaling}, where we use again the short notation $h_a = \iu s_a | \bvec{k}_a| \, u_a$.
However, within the scaling ansatz, the sign of the coefficient of this term is left undetermined, such that the selected configuration may satisfy \emph{either} $\hat{h} \cdot \hat{h} =0$, or $\hat{h} \cdot \hat{h} = 1$ up to some phase (where $\hat{h} = \vec{h} / |\vec{h}|$).
Within the perturbative regime on the other hand, we can identify terms at higher order in the lattice distortion field which leads to the following contribution to the ground state energy
\begin{equation}
    \mathcal{E}^{(4)}[\vec{h}] \sim |\vec{h}|^4 \left( \mathcal{C}_1(\beta) + \mathcal{C}_2(\beta) \left( \hat{h} \cdot \hat{h} \right) \left( \hat{h}^\ast \cdot \hat{h}^\ast \right) \right),
\end{equation}
where $\mathcal{C}_1(\beta)$ and $\mathcal{C}_2(\beta)$ are some constants.
This result is derived in Appendix~\ref{app:massterms} and comes from quadratic- and quartic-order in perturbation theory, considering all possible monopole- and fermion-lattice couplings.
While the sign of $\mathcal{C}_1(\beta)$ is a priori not determined, we find that the most dominant contribution to $\mathcal{C}_2(\beta)$ is positive, from which we infer $\mathcal{C}_2(\beta) > 0$ and thus the selected configuration will satisfy $\hat{h} \cdot \hat{h} = 0$.

\subsubsection{Breaking of emergent $\SO(3)$ symmetry in the ordered phase: Dangerously irrelevant operators}

Under the constraint $\hat{h} \cdot \hat{h} \equiv 0$, it is easily seen that the effective actions (in both the perturbative regime and using the scaling ansatz) possess an $\SOthree_\mathrm{valley}$-symmetry.
A naive analysis might conclude that the spin-Peierls instability is concomitant with a spontaneous breaking of this continuous symmetry group, such that the resulting phase hosts Goldstone modes.

We now argue that this is not the case.
We note that the $\SOthree_\mathrm{valley} \subset \SO(6)$ is an \emph{emergent} symmetry \emph{at the QED$_3$ fixed point}, which follows from the fact that there exist no \emph{relevant} perturbations to the QED$_3$ fixed point which are invariant under the microscopic (UV) symmetries of the system.

Crucially, the physics of the ordered state (i.e. after the system has undergone the spin-Peierls instability) is controlled by some new strong-coupling fixed point FP$^\ast$.
This fixed point need not possess the same emergent symmetry; perturbations allowed by microscopic symmetries can constitute relevant perturbations to FP$^\ast$.
This is the scenario of a \emph{dangerously irrelevant} coupling $\lambda$ \cite{chubu94,senthil04}.
Initializing a renormalization-group (RG) flow in the vicinity of the QED$_3$ (with finite $\lambda,g \neq 0$), the relevant coupling $g$ will increase under the flow to FP$^\ast$, while the dangerously irrelevant $\lambda$ initially decreases.
Upon reaching some crossover scale, $\lambda$ will start to grow as it constitutes a \emph{relevant} perturbation to FP$^\ast$.

Specifically, as pointed out in Ref.~\cite{song19}, the DSL on the triangular lattice admits a three-monopole term
\begin{equation}
    S_\lambda = \lambda \int \du^3 x \left[\Phi_1 \Phi_2 \Phi_3 + \hc \right]
\end{equation}
which is allowed by the microscopic (UV) symmetries (for example, this three-monopole terms has zero lattice momentum since $\bvec{k}_1 + \bvec{k}_2 + \bvec{k}_3 =0$), but clearly breaks the emergent $\SO(6)$ symmetry at the QED$_3$ fixed point.
Note that the three-monopole term has scaling dimension $\Delta_{\Phi\Phi\Phi}\approx4.3$ in the large-$N$ expansion and is therefore assumed to be irrelevant \cite{chester16}.

Considering the coupling of the DSL to displacement fields, any finite $\lambda \neq 0$ induces analogous corresponding anharmonic terms in the effective action,
\begin{equation} \label{eq:e-lambda}
    \mathcal{E}_\lambda[\vec{h}]  \sim \tilde{\lambda} \left[h_1 h_2 h_3 + \hc \right].
\end{equation}
Within perturbation theory about \QED3, one contribution to the effective action of this form arises from $\langle S^3_g S_\lambda \rangle/(\beta V)$.
We stress that, since \eqref{eq:e-lambda} is allowed by the microscopic symmetries of the system, it will also generically arise in \emph{any} microscopic theory of lattice displacements upon going beyond the harmonic approximation, even in the absence of a coupling to the DSL (hence $\tilde{\lambda}$ is not necessarily related to $\lambda$ in $S_\lambda$).

We hypothesize that the dynamics of the field $\vec{h}$ at the strong-coupling fixed point may be described by an O(3) non-linear sigma model \footnote{Upcoming work \cite{upcomingdqcp}. This follows from the non-linear sigma model description of the Dirac spin liquid.}.
Then, any finite $\tilde{\lambda}$ gives rise to a finite mass term for the transverse Goldstone-mode fluctuations. We conclude that $\tilde{\lambda} \neq 0$ explicitly breaks the emergent $\SOthree_\mathrm{valley}$ symmetry of the model \emph{in the ordered phase}.
The resulting energy-minimizing configurations for $\tilde\lambda<0$ are given by
\begin{equation}
    \vec{h}_{\mathrm{GS}}=[u/\sqrt{3}]\left(1, \eu^{2\pi\iu/3}, \eu^{-2\pi\iu/3}\right)^\top.
    \label{eq:h_GS}
\end{equation}
Instead of having a continuous $\SOthree$ manifold of degenerate ground states, there are three degenerate ground states $\eu^{2\pi \iu n/3} \vec{h}_{\mathrm{GS}}$, with $n=0,1,2$. The resultant lattice distortion preserves the $C_3=T_1 C_6^2$ lattice symmetry and has an additional reflection symmetry. In real-space, the pattern is written
\begin{equation}\label{eq:rs12site}
    \bvec{u}(\bvec{x}) = 2 \Im\left( \sum_a s_a \hat{\bvec{k}}_a \, h^a_{\mathrm{GS}} \,\eu^{\iu\bvec{k}_a\cdot \bvec{x}} \right),
\end{equation}
which is highlighted in the left panel of Figure~\ref{fig1} with white arrows.

\begin{figure}
    \centering
\includegraphics[height=0.25\textwidth]{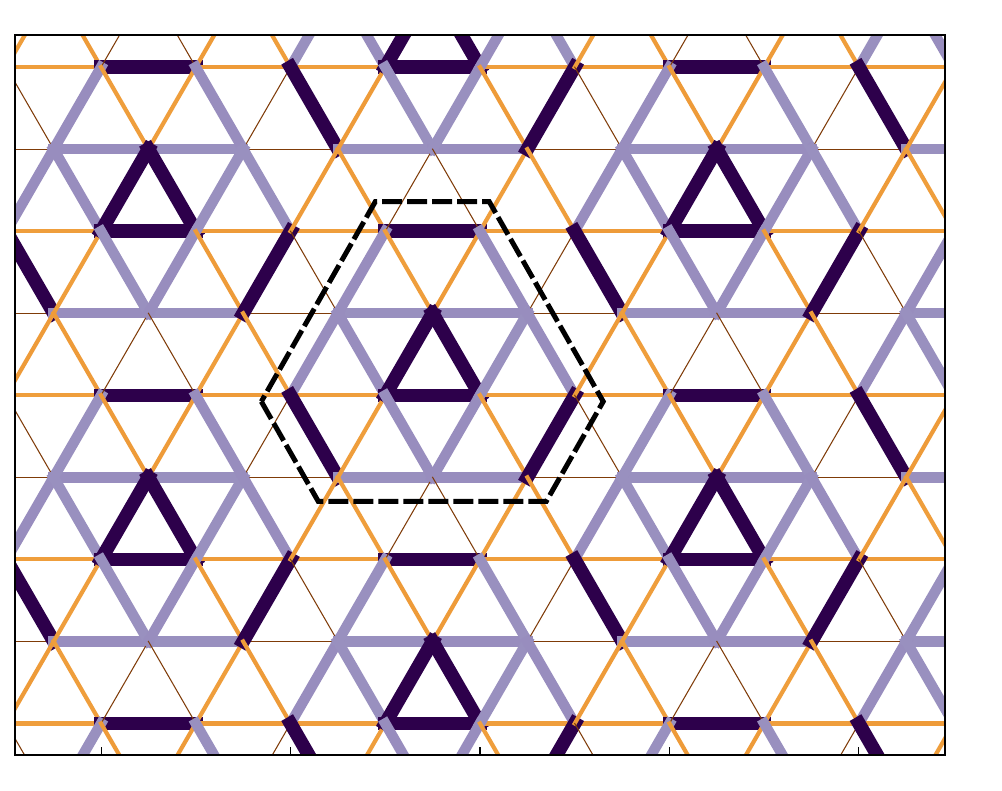}
\includegraphics[height=0.25\textwidth]{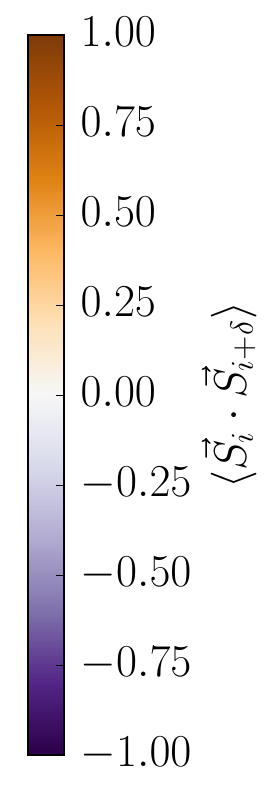}
\caption{\label{fig:dimer_theory}The predicted form of the spin-spin correlation function on dimers $\langle \vec{S}_{i}\cdot\vec{S}_{j}\rangle$ in the presence of a 12-site lattice distortion (defined up to an overall magnitude). Bonds with negative dimer order are plotted with thicker weight; the 12-site order is visible by the strong negative VBS correlations within the unit cells.}
\end{figure}

The spins react to the lattice breaking the translation and $C_6$ rotation by forming a valence bond solid state with the same symmetries.
In the field-theory picture, the choice of $\vec{h} = \vec{h}_{\mathrm{GS}}$ determines the ordering of the monopole excitations as $g\langle\Phi_a\rangle \sim \hat{h}_a$, up to a phase which we choose to be $-1$ (to minimize $g\langle\Phi_a\rangle h_a^*$).
Using the mapping between VBS monopoles and nearest-neighbor spin dimer correlations Eq.~\eqref{eq:vbsmonopolemapping}, we can evaluate the predicted spin-spin correlation function
\begin{equation}
    \langle \vec{S}_{\bvec{x}}\cdot\vec{S}_{\bvec{x}+\bvec{\delta}_a}\rangle
    = s_a \, \Re\left(\langle\Phi_a\rangle\, \eu^{\iu \bvec{k}_a\cdot \bvec{x}}\right).
\end{equation}
This result is correct up to a normalization which symmetry arguments cannot determine. The resultant VBS order is plotted in Fig.~\ref{fig:dimer_theory}. It has the strongest correlations at the boundary of the 12-site unit cell, and has negative dimer correlations on all bonds shortened by the lattice distortion --- this is highlighted in Fig.~\ref{fig1}. 
This pattern is related to the $\sqrt{12}\times\sqrt{12}$ VBS patterns discussed in Refs.~\cite{song19,PhysRevLett.86.1881,PhysRevB.71.224109}; it is potentially generated by resonating, degenerate coverings of the enhanced bonds within the same 12-site unit cell.

\section{Numerical study} \label{sec:numerics}

We now turn to a numerical study of the triangular lattice $J_1$--$J_2$ Heisenberg model, where simulations indicate the presence of a Dirac spin liquid ground state around $J_2/J_1\approx 1/8$ \cite{PhysRevB.42.4800,PhysRevLett.123.207203, Sherman2022, drescher2022dynamical}.
We use the infinite density matrix renormalization group algorithm (DMRG) \cite{White1992, Schollwoeck2011, McCulloch2008, Hauschild2018} to study the lattice model on a cylinder of finite circumference $L_y\equiv L$ \cite{Stoudenmire2012, Gohlke2017} and infinite length $L_x$. This limits the scope of the simulation by the introduction of a finite-size gap to the gapless DSL, but we will find that strong signatures of the DSL response remain, as seen in a previous study of the dynamical structure factor on a cylindrical geometry \cite{Sherman2022, drescher2022dynamical}.

\subsection{Finite geometries}

Our theoretical CFT study predicts that the DSL has a strong energy response to a \emph{static distortion} of the lattice with any of the three momenta $\bvec{k}_a$. In the perturbative `weak coupling' regime in the thermodynamic limit, this was found to be $\sim - g^2 |\bvec{k}_a|^2\beta^{3-2\Delta_\Phi} |\vec{u}|^2$. 
At zero temperature, this quadratic contribution has a divergent prefactor; we expect that even in simulations for a \emph{finite system} (meaning on cylinders with a finite circumference that cuts off our divergences), there will be a strong energy gain from a static lattice distortion with momenta $\bvec{k}_a$. 
The spin-Peierls instability is recovered only in the thermodynamic limit.
For any finite system circumference (smaller than the correlation length), the system's energy will be an \emph{analytic} function in the spin-Peierls coupling $g$, and the change in energy $\Delta \mathcal{E}_{\mathrm{QED}_3}[\vec{u}] =\mathcal{E}_{\mathrm{QED}_3}[\vec{u}]- \mathcal{E}_{\mathrm{QED}_3}[0]$ due to coupling to a distortion field can be obtained using the perturbative expansion introduced in Sec.~\ref{sec:conf-pert-theory},
\begin{equation}
    \Delta \mathcal{E}_{\mathrm{QED}_3}[\vec{u}] = -\lim_{\beta \to \infty} \frac{1}{2 \beta V} \langle S_g^2 \rangle_{\mathrm{QED}_3} + \dots,
\end{equation}
where $\langle S_g^2 \rangle_{\mathrm{QED}_3}$ can be obtained in an analogous manner to \eqref{eq:sg2}, but now performing the space-time integral on the geometry $\lim_{\beta \to \infty} S_\beta^1 \times S_L^2$, such that
\begin{equation}
    \Delta \mathcal{E}_{\mathrm{QED}_3}[\vec{u}] = \tilde{c}_{\Delta_\Phi} g^2 |\bvec{k}_a|^2 \, L^{3-2\Delta_\Phi} |\vec{u}|^2
\end{equation}
where the constant $\tilde{c}_{\Delta_\Phi} = 2 \pi^{3/2} {\Gamma(\Delta_\Phi-1/2)}/[{(3-2\Delta_\Phi)\Gamma(\Delta_\Phi)}]$.
For the small systems that can be simulated numerically, we expect to be in the weak-coupling regime, defined in the new regularization scheme as $g^2 |\bvec{k}_a|^2 L^{3-2\Delta_\Phi} |\vec{u}| \ll 1$.
We conclude that deforming the critical QED$_3$ theory with the relevant coupling $g$ to a distortion field at momenta $\bvec{k}_a$ induces a quadratic term in $g$. Crucially, the power-law scaling of the finite-size dependent prefactor is determined by the scaling dimension of the relevant monopole operators \cite{cardy86}.

In contrast, we expect that distortions at other momenta, e.g., the $\bvec{K}$ or $\bvec{M}$ points, will generically produce a finite response independent of system size. The corresponding energy gain for these patterns may also be calculated within perturbation theory but will produce a finite contribution to $\langle S_g\rangle^2$ due to the finite momentum-transfer (as for the terms with $\bvec{Q}\neq -\bvec{k}_a$ in Eq.~\eqref{eq:before-ope}). 
We aim to provide supporting numerical evidence for our analytical results with a study of the $J_1$--$J_2$ triangular lattice in the DSL phase by demonstrating a response consistent with an instability in the thermodynamic limit, predicted above. 
A response which grows as $L\to\infty$ will eventually lead to the breakdown of the weak-coupling regime for smaller $|\vec{u}|$; in the thermodynamic limit the system will have a non-analytic behavior $|\vec{u}|^\chi$ for all distortions and consequently realize the zero-coupling spin-Peierls instability.
Although behaviour in the thermodynamic limit is impossible to extract from such finite-size simulations, we can use a preliminary finite-size scaling scheme to confirm compatibility with a divergent response for specific distortion patterns.

\subsection{DMRG simulation}

Considering microscopic models, a lattice distortion couples to spins by a nearest-neighbor bond-length-dependent exchange, as shown by a mapping from monopoles to nearest-neighbour spin exchanges in Eq.~\eqref{eq:bdhopping}.
To compare the effect of several different static distortions on the ground state, we take a simple model for the variation of the exchange parameters on the triangular lattice Heisenberg model for different distortion patterns. Namely, we consider a lattice distortion $\bvec{u}_i$ which is periodic in a larger unit cell and then calculate the new nearest- and next-nearest-neighbor lengths $\bvec{r}_{ij}= \bvec{R}_{ij} + \bvec{u}_{ij}$ (where $\bvec{u}_{ij} = \bvec{u}_{i} - \bvec{u}_{j}$).
By assuming that all spin exchange couplings $J_{ij}$ are independent of bond angle, and exponentially dependent on the bond lengths $J_{ij} = J^{(0)} \eu^{-\alpha |\bvec{r}_{ij}|}$ (where $J^{(0)}=J_{1,2}$ for NN/NNN couplings), we derive the leading contribution for small distortion
\begin{equation}
    J_{ij} =  J^{(0)} \left[ 1 - \alpha \,\bvec{u}_{ij}\cdot \bvec{R}_{ij} / |\bvec{R}_{ij}| \right].
\end{equation}
Hereon, we absorb the dimensionful prefactor $\alpha$ into the distortion $\bvec{\delta}_{ij}=\alpha\,\bvec{u}_{ij}$, and think of distortions where $|\bvec{\delta}_{ij}| \ll 1$. For the $C_3$ symmetric pattern, the bond-strengths are highlighted in the left panel of Fig.~\ref{fig1}. For all simulations presented in this work, we considered a  ratio of (undistorted) nearest to next-nearest neighbour coupling of $\frac{J_2}{J_1} = 1/8$, which lies within the regime of the putative Dirac quantum spin liquid phase \cite{Hu2019, Sherman2022, drescher2022dynamical}.

\begin{figure}
\centering
\includegraphics{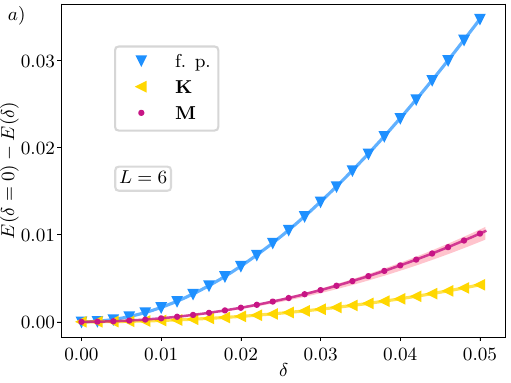}
\includegraphics[scale=1]{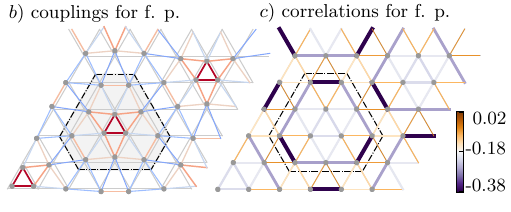}
\caption{(a) Energy response relative to the ground state energy for the different distortion patterns generated by the momenta ${\bvec K} = -{\bvec K}_3$, the averaged response for the ${\bvec M}$-points (${\bvec M}_i$, $i=1,2,3$) and the `full pattern' (f. p.) of the 12-site unit cell as defined in Eq.~\eqref{eq:rs12site}. The shaded region of the ${\bvec M}$-curve indicates the range of different orientations of the pattern on the cylinder. A detailed discussion of the convergence is available in Fig.~\ref{fig:convergence_full_pattern_all_M} in the appendix; we used DMRG simulations with bond dimensions up to $\chi = 4000$ on an $L=6$-cylinder. A depiction of the real-space distortions of the couplings is at hand in Figs.~\ref{fig:Numerical_appendix_patterns}--\ref{fig:distortion_patterns_energy_response}. (b) The distorted nearest-neighbor bonds of the triangular lattice Heisenberg model with the full 12-site distortion pattern; red/blue colors indicating enhanced/weakened bonds. (c) The spin-spin correlation function on dimers $\langle \vec{S}_{i}\cdot\vec{S}_{j}\rangle$ in the presence of the same distortion $\delta=0.05$, showing enhancement at the boundaries of the 12-site unit cell. The plotted color scale is centred on $-0.18$, the calculated uniform value of the spin-spin correlation in the DSL phase on the undistorted lattice. Purple/orange bonds indicate more negative/positive correlations than the DSL.
}
\label{fig:energygain}
\end{figure}

\subsubsection{Energy response to distortion}
\label{sec:energy_response_to_distortion}
We will now compare the response of four patterns: the three-site distortion with momentum $-\bvec{K}_3$, the two-site distortions for momenta $\bvec{M}_i$ ($i=1,2,3$), the six-site distortion $\bvec{k}_3 = -\bvec{K}_3/2$, and the 12-site pattern comprised of all $\bvec{k}_{a=1,2,3}$ [this distortion pattern is plotted in Fig.~\ref{fig:energygain}(b)]. We predict that this $C_3$ symmetric pattern, defined in real-space by Eq.~\eqref{eq:rs12site}, will show the strongest response.
The patterns given in terms of a single momentum eigenstate $\bvec{Q} = -\bvec{K}_3, \bvec{M}_i, \bvec{k}_3$ ($i=1,2,3$) are given simply by 
$\bvec{u}_i = u (\cos[\bvec{Q}\cdot \bvec{x}_i + \theta], \sin [\bvec{Q}\cdot \bvec{x}_i+\theta])$, where $\theta \in [0, 2\pi]$ is an arbitrary but fixed phase. The detailed phases used are specified in Appendix~\ref{appendix:numerical_simulations}.
In all cases, the states can be written as either one or a sum of three momentum eigenstates satisfying the normalization $\sum_{\bvec{q}} |\bvec{u}_{\bvec{q}}|^2 = u^2$.
The energy cost of all patterns is defined through Eq.~\eqref{eq:distcost}, giving $\mathcal{H}[u] = \mathcal{K}_{\bvec{Q}} u^2$ with
\begin{equation}
    \mathcal{K}_{\bvec{M}} = \frac{8 K}{3\mathsf{a}^2},\,\,
    \mathcal{K}_{\bvec{K}} = \frac{5 K}{3\mathsf{a}^2},\,\,
    \kappa = \mathcal{K}_{\bvec{k}_a} = \frac{(5-2\sqrt{3}) K}{3\mathsf{a}^2}.
\end{equation}
The patterns with larger unit cells have a generally smaller energy cost for the same momentum-space distortion magnitude $u^2$, meaning the $\bvec{k}_a$ patterns have the lowest potential energy cost of the patterns we will compare. Note that in simulations of a distorted Heisenberg model, we calculate the spin energy gain in terms of the dimensionless distortion parameter $\delta = \alpha u$, where the constant $\alpha$ is equal for all patterns. Whether or not a finite-size lattice will spontaneously and statically distort with a given pattern, is determined by the relative magnitudes of the energy gain and the potential energy cost $\mathcal{H}[\delta] =( \mathcal{K}_{\bvec{Q}}/\alpha^2 - A_{\bvec{Q}}) \, \delta^2$, where $A_{\bvec{Q}}$ is the coefficient of the energy gain of the spins, to be computed numerically.

To model the lattice distortion, we stabilize the spin-disordered ground state on a translationally invariant lattice (we take the $J_2 = J_1/8$ parameter value in the spin-liquid regime).
We simulate a cylindrical geometry infinite in the $x$-direction by repeating a $L_x=3,6$ unit cell (chosen to be compatible with the distortion pattern). We consider first a cylinder with circumference $L_y=6$ using the $YC6$ boundary conditions \cite{Zhu2015, Hu2019} (see Appendix~\ref{appendix:numerical_simulations} for details). 
We then introduce a small distortion of the lattice according to one of the patterns, distort the NN and NNN bonds accordingly, and then use DMRG to find the ground state on the new lattice. We use the spin-disordered ground state as the starting run and make no assumptions about the resultant spin state.
Increasing the distortion parameter $\delta$ proceeds by increasing the lattice distortion and calculating the resultant ground state with its energy.
This process is repeated for increasing bond dimension $\chi$ (up to $4000$) in order to ensure the resultant energy differences are well converged.
A detailed description of this method, as well as figures showing all distortion patterns investigated, is provided in Appendix~\ref{appendix:numerical_simulations}.

A comparison of the energy gain of the spins $-\Delta \mathcal{E}[\delta] =\mathcal{E}[0] - \mathcal{E}[\delta] $ in the presence of three lattice distortions on the $L_y=6$ cylinder are shown in Figure~\ref{fig:energygain}. We model the energy gain as $-\Delta \mathcal{E}[\delta]=A_{\bvec{Q}}\delta^\alpha$ and use this form to fit the data.
All curves are consistent with a quadratic response at small distortions, and we can see that the 12-site unit cell pattern has a much stronger response amplitude. The coupling of the DSL to a lattice distortion commensurate with the monopole excitations leads to the strongest response. We also note that the symmetric combination of all three momenta $\bvec{k}_a$ leads to a stronger response than $\bvec{k}_3$ individually, as predicted in Section~\ref{sec:symdistfield}. 
An analysis of the convergence of the DMRG calculations, as well as the results of the fitting performed with logarithmic axes is provided in Appendix~\ref{appendix:numerical_simulations}. We conclude that at this bond dimension, the change in energy due to distortion is well converged on $L_y=6$ geometries.

\subsubsection{VBS order}
Now we turn to the VBS order on the deformed lattices. In Fig.~\ref{fig:energygain}(c), the nearest-neighbor spin-spin correlations $\langle \vec{S}\cdot\vec{S}\rangle$ of the symmetric 12-site patterns are plotted and compared to the uniform correlations of the DSL ground state in the undistorted model.
The correlations shown have been obtained for bond dimension $\chi = 2000$ after adiabatically increasing the distortion $\delta$. The undistorted ground state had been optimized using the odd-sector method for the same bond dimension (see Appendix~\ref{appendix:numerical_simulations} for details).
It matches well with our prediction of the condensation of three monopoles $\langle\Phi_a\rangle \sim \eu^{2\pi\iu \, a / 3}$ (cf. Fig.~\ref{fig:dimer_theory}), showing strong negative VBS weight (relative to the DSL) on the boundaries of the 12-site unit cell (particularly on the shorter sides, which we observe are stronger than the central triangle). 
Importantly, the more negative VBS weights are inside the unit cell along bonds that are shortened by the distortion, and we additionally see a more positive weight between the unit cells along lengthened bonds. 
The large variation of bond order (up to range $\approx 0.4$) in the VBS state is \emph{significantly} larger than the almost uniform DSL state, with correlations $-0.181$
and variations of order $10^{-3}$. This variation in dimer correlations serves as a VBS order parameter and is a clear sign here of the transition away from the DSL.
We have hence seen that the 12-site lattice distortion leads to a \emph{strong energetic gain} and \emph{transition to a short-range correlated phase}, which is consistent with a DSL-VBS transition on a finite-circumference cylinder.

\begin{figure}
\centering
\includegraphics{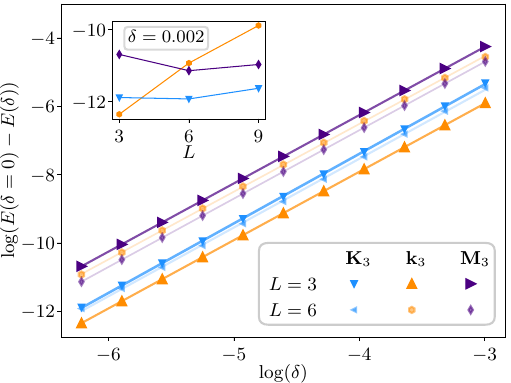}
\caption{Energy response relative to the ground state energy for the different distortion patterns generated by the momenta ${\bvec K}_3$, ${\bvec k}_3$ and ${\bvec M}_3$ for $L=6$ and $L=3$ (smaller, lighter markers are $L=6$). The ground state $L=6$ has been found by the odd-sector method and is well converged (see Appendix~\ref{appendix:numerical_simulations}). The exponents $\alpha$ can be extracted from a linear fit, all values are consistent with the integer value $2$. We plot the data at the highest available bond dimension. The inset shows $\log(E(\delta=0)-E(\delta=0.002))$ for a cylinder circumference of up to $L=9$. The hexagons represent ${\bvec k}_3$, the triangles ${\bvec K}_3$ and the diamonds ${\bvec M}_3$.}
\label{fig:scaling_analysis}
\end{figure}

\subsubsection{System-size dependence}

Next, we test the prediction of the weak-coupling CFT calculation and connect to the thermodynamic limit by performing finite-size scaling.
We will show that the magnitude of the energy response as a function of system size $A_{\bvec{Q}}(L)$ is only non-trivial for the static distortions that couple to monopoles.

We consider finite-size systems with circumference $L_y=3,6,9$; since the symmetric pattern cannot fit on these new geometries, we are restricted to only studying the momentum-eigenstate patterns $\bvec{Q} = \bvec{M}_3, \bvec{K}_3, \bvec{k}_3$.
These patterns all fit in a six-site unit cell, which we can fit on the $\mathrm{YC}N$ geometry for $N$ a multiple of 3.

The responses for the three patterns with bond dimension up to $\chi=7000$ are plotted in Figure~\ref{fig:scaling_analysis}.
The responses $\Delta E(\delta)$ for $L=3,6$ are well converged for all $\delta$ and well described by a quadratic response $\alpha=2$; for $L=9$, we are able to converge one point $\delta=0.02$ for the three patterns. Due to this numerical limitation, we cannot compare the amplitudes obtained from fitting for all system sizes, so instead we focus on comparing the energy gain $\Delta E(\delta=0.002)$ in the inset as a function of system size.
We find that only the monopole pattern with $\bvec{k}_3$ shows significant system-size dependence, with its amplitude increasing strongly. In contrast, the $\bvec{K}_3$ response amplitude does not significantly increase, and the $\bvec{M}_3$ response even slightly reduces in amplitude.
This difference of behavior under variation of cylinder circumference seems to confirm that the system is unstable to only lattice distortions with momenta $\bvec{k}_a$ in the thermodynamic limit.
We conclude that the numerical simulations, even though performed on small cylinder circumferences, are compatible with our prediction of a zero-coupling instability in the thermodynamic limit.

\section{Dynamical phonons} \label{sec:dyn-phon}

In the derivation of the weak-coupling spin-Peierls instability, we made the crucial assumption that the lattice displacements can be treated as a classical background field without any intrinsic dynamics.
This \emph{adiabatic approximation} is justified if the magnetic energy scales (such as bare microscopic exchange couplings) are much larger than those of the lattice displacements, such that the lattice dynamics occurs on much longer time scales compared to the spin subsystem.
In the case of the one-dimensional spin-Peierls transition, the treatment of dynamical phonons removes the zero-coupling instability and produces a stable spin-liquid regime \cite{citro05}.

In this section, we go beyond the adiabatic assumption for the DSL.
To this end, we focus on the \emph{relevant} interaction between valley-triplet monopoles and lattice distortions at wavevectors $\bvec{k}_a$ with $a=1,2,3$ and now promote the corresponding Fourier components to dynamical fields $\bvec{u}_{\bvec{k}_a} \to \bvec{u}_{\bvec{k}_a}(\tau,\bvec{x})$.

\subsection{Integrating out dynamical phonons}

We add dynamics for the long-wavelength fluctuations of the $\bvec{k}_a$-displacement modes via a (heuristic) kinetic energy term, $\mathcal{S}_\mathrm{kin} = \rho \sum_a |\partial_\tau u_a(\bvec{x},\tau)|^2$, where we directly work in terms of the longitudinal components $u_a = \hat{\bvec{k}}_a \cdot \bvec{u}_{\bvec{k}_a}$ and $\rho$ corresponds to a microscopic mass density of phonon degrees of freedom. The lattice stiffness is given as before by $\kappa \sim K / \mathsf{a}^2$, but is now taken to be independent of momentum.
Here, $u_a(\tau,\bvec{x})$ is to be understood as a complex scalar field, since the Fourier components $\bvec{u}_{\bvec{k}_a} = \bvec{u}_{-\bvec{k}_a}^\ast $.
The full (dynamical) action then reads
\begin{equation} \label{eq:s-ph-dyn}
    S_\mathrm{ph}[u_a] = \int\! \du \tau \! \int \!\du^2 \bvec{x} \left[ \rho | \partial_\tau u_a(\tau,\bvec{x})|^2 + \kappa |u_a(\tau,\bvec{x})|^2 \right].
\end{equation}
Note that for simplicity we have not included gradient terms which would induce a long-wavelength dispersion for the $u_a(\bvec{x},\tau)$-fields, since $S_\mathrm{ph}$ already includes an explicit energy gap $\omega_0 \equiv \sqrt{\kappa/\rho}$ for the phonons.
Writing $u_a(\tau) = (2 \pi)^{-1} \int \du \omega\, \eu^{-\iu \omega\tau} u_a(\omega)$ and Fourier-transforming $G(\omega) = (\rho \omega^2 + \kappa)^{-1}$, we find the phonon Green's function from Eq.~\eqref{eq:s-ph-dyn} as
\begin{equation}
    \langle u_a (x) u_b^\ast(y) \rangle_\mathrm{ph} = \delta_{a,b} G(x-y) = \delta_{a,b} \frac{\delta^{(2)}(\bvec{x}-\bvec{y})}{2 \rho \omega_0} \eu^{-\omega_0 |\tau_x-\tau_y|}, \label{eq:g-exp}
\end{equation}
again with the short notation $x=(\tau, \bvec{x})$.
The Green's function $G(x-y)$ simply corresponds to a local imaginary-time Green's function of a harmonic oscillator.

The full action $S$ of the system then reads
\begin{equation}
    S[u_a(\tau, \bvec{x})]  = S_{\mathrm{ph}}[u_a(\tau,\bvec{x})] + S_{\mathrm{QED}_3} + S_g,
\end{equation}
where $S_{\mathrm{QED}_3}$ is the fixed-point action for QED$_3$, and $S_g = \int \du \tau \du^2 \bvec{x}\, \mathcal{H}_g$ is the coupling between monopoles and (now dynamical) phonons, see Eq.~\eqref{eq:h-g-explicit}.
Crucially, as long as the phonons remain gapped $\omega_0 > 0$, these degrees of freedom can be integrated out exactly (owing to their quadratic, non-interacting nature) to obtain an effective action for the \QED3 degrees of freedom.
Explicitly, using the equation of motion
\begin{equation}
    u_a(x) = \big( \iu |\bvec{k}_a| g s_a \big) \int \du^3 y \, G(x-y) \Phi_a(y)
\end{equation}
we find the system's full action $S = S_{\mathrm{QED}_3} + S_{\Phi \Phi}$ where integrating out the dynamical phonon degrees of freedom induces an effective interaction among the monopole operators,
\begin{equation} \label{eq:s-phiphi}
    S_{\Phi\Phi} = - g^2 \sum_a |\bvec{k}_a|^2 \int \du^3 x \,\du^3 y \ \Phi_a^\dagger(x) G(x-y) \Phi_a(y).
\end{equation}
We note that the monopole-monopole interaction in \eqref{eq:s-phiphi} is local in space but \emph{retarded} in time.

\subsection{Phase transition at finite frequencies} \label{sec:dynamic-scaling}

In the adiabatic limit $\omega_0=0$ the monopole-monopole interaction in \eqref{eq:s-phiphi} becomes non-local in time. This implies that modes become correlated at large (temporal) separations, leading to an ordering instability at infinitesimal couplings as per the results of Sec.~\ref{sec:instability}.
Continuing to finite frequencies $\omega_0 > 0$, one may be tempted to perform a mean-field decoupling of the monopole-monopole interaction in $S_{\Phi \Phi}$. This would capture an ordering instability by some \emph{finite} (and time-independent) monopole expectation value \cite{luo22}, $\Phi(x) \to \langle \Phi \rangle_\mathrm{MF}(x) \neq 0$.
However, an explicit perturbative mean-field calculation (again employing an explicit IR regularization, e.g. by working at finite temperature) reveals a criticality condition identical to the static case obtained in Sec.~\ref{sec:sp-instab}, which \emph{crucially} does not depend on $\omega_0$.
This is understood to be an artifact of mean-field theory which neglects the retardation of the monopole-monopole interaction in $S_{\Phi\Phi}$, and thus cannot capture any effects due to having non-zero phonon frequency $\omega_0$.
This may also be seen explicitly: After decoupling, (schematic notation) $\Phi^\dagger \Phi \to \langle \Phi^\dagger \rangle_\mathrm{MF} \Phi + \Phi^\dagger \langle \Phi \rangle_\mathrm{MF} - \langle \Phi^\dagger \rangle_\mathrm{MF} \langle \Phi \rangle_\mathrm{MF}$, with time-independent $\langle \Phi \rangle_\mathrm{MF}$, the effective action becomes equivalent to the classical action $\mathcal{H}_\mathrm{ph} + \mathcal{H}_g[\bvec{u}] + \mathcal{S}_{\mathrm{QED}_3}$ considered in Sec.~\ref{sec:instability}, if one relates the monopole mean-field expectation value to an effective displacement field.

In the opposite, antiadiabatic, limit $\omega_0\rightarrow\infty$ (with $\kappa$ constant), the Green's function becomes purely local, $G(x-y)=(1/2\rho)\,\delta^{(3)}(x-y)$.
Note that in this limit, the monopole-monopole interaction becomes singular due to the OPE $\lim_{\epsilon->0}\Phi^\dagger(\epsilon+x) \Phi(x) \sim \mathds{1} / \epsilon^{2\Delta_\Phi}$. To regularize this, one may work with a finite UV cutoff $ a \leq |x-y| $ that enforces a lower bound on the monopole-monopole separation.
Then, upon taking $\omega_0 \to \infty$ at some \emph{fixed} cutoff, the interaction effectively vanishes and the DSL remains stable. Considering both limits, it becomes clear that there must exist some transition at some intermediate coupling $g$ determined by $\omega_0$.

To obtain the critical interaction strength as a function of $\omega_0$, we perform a scaling analysis using the exponential form of $G(\tau_x-\tau_y)$ in Eq.~\eqref{eq:g-exp}: For separations $|\tau_x - \tau_y| \gtrsim 1/\omega_0$, the interaction is exponentially suppressed, thereby setting a critical cutoff scale $a_c \sim 1/ \omega_0$. Starting with some cutoff $a < a_c$, we can increase the cutoff up to $a_c$ where the interaction becomes suppressed. Rescaling all coordinates $x = x' /a_c$ (with $x'$ now dimensionless) reveals the critical (dimensionless) interaction scale $1 \sim g^2 \omega^{2\Delta_\Phi -3}/(\mathsf{a}^2 \kappa)$, or equivalently $g_c^2 \sim \kappa \mathsf{a}^2 \omega_0^{3-2\Delta}$ (where we use that $|\bvec{k}| \sim 1 /\mathsf{a}$ with $\mathsf{a}$ being the lattice spacing).

\subsection{Perturbation theory}

We now investigate the stability of the DSL in the presence of the interaction $S_{\Phi\Phi}$ within perturbation theory.
Considering a perturbative expansion of local observables, the expansion's breakdown is taken to signal an instability of the DSL.
We employ the path-integral formalism, where the expectation value of some observable $\mathcal{X}$ is written as $\langle \mathcal{X} \rangle = \mathcal{Z}^{-1} \int \mathcal{D}[\{\mathcal{O}_\mathrm{CFT}\}] \, \mathcal{X} \eu^{-S}$ where $S$ is the system's interacting action and $\mathcal{Z}$ its partition function without operator insertion, as above.

At the QED$_3$ fixed point (i.e. zeroth order perturbation theory), all one-point functions vanish at zero temperature $\beta \to \infty$ in the thermodynamic limit by conformal symmetry.
We thus turn to the monopole two-point function at large distances $|d| = |(\bvec{d},\tau_d)| \to \infty$.
Expanding the Boltzmann weight $\eu^{-S_{\Phi \Phi}}$ in the path integral to first order, one finds the perturbative expansion
\begin{multline} \label{eq:phi-phi-g2}
    \langle \Phi^\dagger_a(d) \Phi_b(0) \rangle_{g^2} = \frac{\delta_{ab}}{|d|^{2\Delta_\Phi}} \\
    - \left( \langle \Phi^\dagger_a(d) \Phi_b(0) S_{\Phi \Phi} \rangle_{\mathrm{QED}_3} - \frac{\delta_{ab}}{|d|^{2\Delta_\Phi}} \langle S_{\Phi \Phi} \rangle_{\mathrm{QED}_3}  \right) + \dots,
\end{multline}
where we have used Eq.~\eqref{eq:cft-correl} for two-point functions evaluated at the CFT fixed point.
The first term in the parenthesis is written explicitly as
\begin{widetext}
\begin{equation} \label{eq:phi-phi-S-long}
    \langle \Phi^\dagger_a(d) \Phi_b(0) S_{\Phi \Phi} \rangle_{\mathrm{QED}_3} = - \sum_c \frac{g^2 |\bvec{k}_c|^2}{2\rho \omega_0} \int \du^3 x \du^3 y \left[ \delta^{(2)}(\bvec{x}-\bvec{y}) \eu^{-\omega_0 |\tau_x - \tau_y|} \langle \Phi^\dagger_a(d) \Phi_b(0) \Phi^\dagger_c(x) \Phi_c(y) \rangle_{\mathrm{QED}_3} \right].
\end{equation}
\end{widetext}
Note that while 2- and 3-pt. functions of primaries are exactly determined by conformal symmetry, no closed forms are available for 4-pt. functions, making the evaluation of Eq.~\eqref{eq:phi-phi-g2} challenging.

While Wick's theorem does not hold in interacting CFT, we can successively apply operator product expansions (OPE) to decompose the 4-pt. function in \eqref{eq:phi-phi-S-long}.
If the full (convergent) OPE is used, the choice of OPE channels is inconsequential (indeed, demanding OPE associativity lies at the heart of CFT bootstrap studies \cite{poryvi19}).
However, for analytical tractability, we will use the asymptotic forms of OPE [see also Eq.~\eqref{eq:def-ope}].
A particular OPE channel then corresponds to a particular configuration of operator insertions which yield the most divergent contribution in the particular channel if they become `close'.

Considering Eq.~\eqref{eq:phi-phi-S-long}, we will focus on the leading monopole-antimonopole OPE channel (schematically $\Phi^\dagger \times \Phi = \mathds{1}$) corresponding to $x \to 0$ and $y \to d$, with $x$ and $y$ sufficiently far apart, which means that we may assume $0 < |x| \ll |y|$. Explicitly, the first-order OPE in this channel then reads
\begin{equation} \label{eq:dom-channel}
    \langle \Phi^\dagger_a(d) \Phi_b(0) \Phi^\dagger_c(x) \Phi_c(y) \rangle_{\mathrm{QED}_3} \sim \delta_{ac} \delta_{bc} \frac{1}{|d-y|^{2\Delta_\Phi} |x|^{2\Delta_\Phi}}.
\end{equation}
We expect this channel to give the most dominant contribution:
The other monopole-antimonopole channel corresponding to $d\to 0$, $x\to y$ is asymptotically equivalent to the disconnected contribution (second term in the parenthesis) in Eq.~\eqref{eq:phi-phi-S-long}, and is thus expected to only weakly contribute to $\langle \Phi^\dagger_a(d) \Phi_b(0) \rangle_{g^2}$.

We use \eqref{eq:dom-channel} in \eqref{eq:phi-phi-S-long}.
As $|d| \to \infty$, we can take $|d-y|^{-2\Delta_\Phi} \approx |d|^{-2\Delta_\Phi}$ and subsequently perform the $\bvec{y}$-integration, yielding
\begin{multline}
    \langle \Phi^\dagger_a(d) \Phi_b(0) S_{\Phi \Phi} \rangle_{\mathrm{QED}_3} \approx \\
    - \frac{\delta_{ab}}{|d|^{2\Delta_\Phi}} \frac{g^2 |\bvec{k}_a|^2}{2\rho \omega_0} \int \du^3 x \int \du \tau_y \,  \frac{\eu^{-\omega_0 |\tau_x - \tau_y|} }{(|\bvec{x}|^2 + \tau_x^2)^{\Delta_\Phi}}.
\end{multline}
We work on a cylindrical geometry $S_\beta^1 \times \mathbb{R}^2 \overset{\beta \to \infty}{\to} \mathbb{R}^3$.
The spatial integration is performed in polar coordinates with $0 < |\bvec{x}| < L \to \infty$,
\begin{multline}
    |d|^{2\Delta_\Phi} \langle \Phi^\dagger_a(d) \Phi_b(0) S_{\Phi \Phi} \rangle_{\mathrm{QED}_3} \approx \\
     - \delta_{ab} \frac{ \pi g^2 |\bvec{k}_a|^2}{2 (\Delta_\Phi-1) \rho \omega_0} \int \du \tau_x \du \tau_y \, \frac{\eu^{-\omega_0 |\tau_x -\tau_y|}}{|\tau_x|^{2\Delta_\Phi - 2}}.
\end{multline}
Note that due to the $\delta$-distribution for the spatial components in \eqref{eq:phi-phi-S-long}, only configurations with $\bvec{x} = \bvec{y}$ give a finite contribution.
Thus, our previous assumption of $|x| \ll |y|$ implies for the temporal coordinates $|\tau_x| \ll |\tau_y|$.
With $|\tau_x - \tau_y| \approx |\tau_y|$, the remaining integrals are obtained as
\begin{multline}
    \lim_{\beta \to \infty} \int_0^\beta \du \tau_y \int_0^{|\tau_y|} \du \tau_x \, \frac{\eu^{-\omega_0 | \tau_y|}}{|\tau_x|^{2\Delta_\Phi-2}} \approx
     \frac{1}{3-2\Delta_\Phi} \frac{\Gamma(4-2\Delta_\Phi)}{\omega_0^{4-2\Delta_\Phi}},
\end{multline}
which exists for $\Delta_\Phi < 2$, which is within the range of assumed scaling dimensions.
Hence, one finds
\begin{equation}
    |d|^{2\Delta_\Phi} \langle \Phi^\dagger_a(d) \Phi_b(0) S_{\Phi \Phi} \rangle_{\mathrm{QED}_3} \approx - \delta_{ab} \frac{c_{\Delta_\Phi} \Gamma(4-2 \Delta_\Phi)}{4 \rho \omega_0^2} \frac{g^2 | \bvec{k}_a|^2}{\omega_0^{3-2 \Delta_\Phi}}
\end{equation}
Generically, perturbation theory breaks down when the $(n+1)$-term in the expansion is no longer small compared to the $n$-th term.
Here, we compare the first-order correction to the bare correlator,
\begin{equation}
    \frac{\delta^{(1)} \langle \Phi_a^\dagger(d) \Phi_a(0) \rangle_{g^2}}{\langle \Phi_a^\dagger(d) \Phi_a(0)\rangle_{\mathrm{QED}_3}} \approx \frac{1}{4} c_{\Delta_\Phi} \Gamma(4-2\Delta_\Phi) \frac{g^2 |\bvec{k}_a|^2}{\kappa \omega_0^{3-2\Delta_\Phi}}.
\end{equation}
Perturbation theory breaks down if this ratio is of order 1.
This leads to a scaling relation for the critical coupling, given by
\begin{equation} \label{eq:phonon-gc}
    g^2_c \sim K \omega_0^{3-2\Delta_\Phi}.
\end{equation}
In the thermodynamic limit at zero temperature, the finite phonon frequency $\omega_0 \neq 0$ hence prevents a weak-coupling instability and rather determines a critical $g_c$ as a function of $\omega_0$. The critical $g_c$ matches precisely the result of the scaling analysis in Sec.~\ref{sec:dynamic-scaling}.

\subsection{Monopole-induced phonon softening and Kohn-type anomalies}

While in the previous subsections we have integrated out the dynamical phonons and analyzed the resulting monopole-monopole interactions in the Dirac spin liquid, we here take the opposite route and investigate how the phonon-monopole coupling is manifested in spectral properties of the phonon.
In particular, we conclude that there exists a Kohn-like anomaly \cite{PhysRevLett.2.393,PhysRevB.9.2911} in the phonon dispersion, i.e. at (and close to) the wavevectors $\bvec{K}_a/2$ there will be a characteristic softening of the otherwise gapped (away from $\bvec{q}\equiv 0$) phonon. This can be understood in analogoy to the Kohn anomaly due to the divergent particle-hole susceptibility at $2k_\mathrm{F}$ in 1+1-dim. systems \cite{PhysRevLett.2.393,PhysRevB.9.2911}.

Working at second order perturbation theory, we find that \eqref{eq:h-g-explicit} yields a correction to the propagator of the phonon mode with lattice momentum $\bvec{K}_a/2$ [cf. Eq.~\eqref{eq:s-ph-dyn}], given by
\begin{equation} \label{eq:g-renorm}
    G^{-1}_a(\omega) = \rho \omega^2 + \kappa - g^2 |\bvec{k}_a|^2 \langle \Phi^\dagger_a \Phi_a \rangle_{\mathrm{QED}_3}(\omega).
\end{equation}
Continuing to real frequencies, this implies that the phonon quasiparticle dispersion $\omega_a$ becomes renormalized and is determined by solutions to the implicit equation
\begin{equation} \label{eq:disp-implicit}
    \omega^2 = \omega_0^2 - {g^2 |\bvec{k}_a|^2}{\rho}^{-1} \chi_a'(\omega,T),
\end{equation}
where $\chi'$ is the real part of the (in general $T$-dependent) VBS susceptibility $\chi'_a(\omega,T) \equiv \Re \chi_\mathrm{VBS}(\omega,\bvec{k}_a+\delta \bvec{q},T)|_{\delta\bvec{q} = 0}$, which for momenta close to $\bvec{k}_a$ is given by singular VBS monopole-monopole correlations \cite{herm05}.

For temperatures $T \gg \omega$, scaling arguments imply that to leading order $\chi_a'(\omega,T) \sim T^{-(3-2\Delta_\Phi)} \left(c + \mathcal{O}(\omega/T)\right)$ with some constant $c$ \cite{witz15}, and from \eqref{eq:disp-implicit} it follows that the phonon frequency (pole of $G_a(\omega)$) is shifted downwards.
Importantly, this provides an experimentally accessible signature of the DSL via monopole-phonon couplings, in particular also in parameter regimes where the spin liquid is expected to be stable.
The softening of the phonon will exhibit a power-law scaling with temperature with a  (universal) exponent that depends on the monopole operator dimension as a characteristic signature of the Dirac spin liquid.

We can find an explicit solution to the equation \eqref{eq:disp-implicit} by considering $\omega \to 0$ and $T > 0$.
In this case, above scaling form for $\chi_a'$ becomes exact, yielding $0= \omega_0^2 - g^2 | \bvec{k}_a|^2 \rho^{-1} \chi'_a(0,T)$.
This is solved when $1 \sim g^2 |\bvec{k}_a|^2 \kappa^{-1} T^{-(3-2\Delta_\Phi)}$, which precisely recovers the critical scaling of the spin-Peierls temperature $T_\mathrm{SP}$ as a function of $g$ in Eq.~\eqref{eq:tsp}.
At this critical temperature, the softened phonon hits zero energy and condenses, giving rise to a finite static lattice distortion as discussed in Sec.~\ref{sec:instability}.

Conversely, we can analyze the zero-temperature $T\to 0$ limit in \eqref{eq:disp-implicit} working perturbatively: If the second term on the right-hand side of \eqref{eq:disp-implicit} is much smaller than first term, one can perturbatively solve the implicit equation by iteratively substituting (the square-root of) the left-hand side for $\omega$ on the right-hand side, generating an order-by-order expansion.
To leading order, we thus obtain
\begin{equation} \label{eq:implicit-sol-freq}
    \omega^2 \approx \omega_0^2 - g^2 |\bvec{k}_a |^2 \rho^{-1} \chi_a'(\omega_0,0) + \dots.
\end{equation}
Using $\chi_a'(\omega,0) \sim \omega^{-(3-2\Delta_\Phi)}$, again fixed by scaling \cite{herm05,witz15}, one obtains that the phonon dispersion is renormalized down to zero energy $\omega = 0$ when $1 \sim g^2 |\bvec{k}_a|^2 \kappa^{-1} \omega_0^{-(3-2\Delta_\Phi)}$, which coincides with the parameter regime where the perturbative treatment of \eqref{eq:disp-implicit} breaks down, and yields the scaling law \eqref{eq:phonon-gc} for the critical spin-Peierls coupling at finite frequency and zero-temperature.

While above analysis was focussed on the dynamics of the phonon modes at the relevant wavevectors $\bvec{k}_a$, one can extend the interacting phonon propagator $G_a(\omega) = G(\omega,\bvec{k}_a)$ in Eq.~\eqref{eq:g-renorm} to momenta $\bvec{k}_a + \delta \bvec{q}$ around $\bvec{k}_a$ (with $|\delta \bvec{q}| \ll |\delta \bvec{k}_a|$) given a microscopic model for the bare phonon dispersion $\omega_0(\bvec{q})$, as e.g. in \eqref{eq:distcost}.
With the zero-temperature form of the monopole susceptibility $\chi(\omega,\bvec{k}_a+\delta \bvec{q}) = [c^2|\delta \bvec{q}|^2-\omega^2]^{-(3/2 - \Delta_\Phi)}$, the phonon spectral function (at second order perturbation theory) can be calculated as shown in Appendix~\ref{sec:phononspecfn}.
Turning to finite temperatures, we note that the monopole susceptibility can be written in terms of some unknown scaling function, but in contrast to (1+1)-dimensional quantum critical theories, there does not exist a conformal mapping which would allow us to extract the dynamical response at finite temperatures.
Gaining an in-depth understanding of finite-temperature dynamical correlations of the DSL is a highly interesting task for future theoretical and computational studies \cite{witz15,lucas17}.

\section{Conclusion \& Outlook}\label{sec:conclusion}

\subsection{Summary}

We have studied the effective theory of the U(1) Dirac spin liquid in the presence of lattice distortions.
The effective low-energy description of spinons coupled to a dynamical gauge field is \QED3 with $N_f=4$ fermions which is believed to flow to a strongly coupled conformal fixed point in the continuum. 
The compactness of the dynamical gauge field allows for monopoles as instanton tunneling events which insert a $2\pi$ flux of the emergent gauge field.
While on the triangular lattice these operators are relevant, they carry non-trivial quantum numbers under the microscopic UV lattice symmetry, and the isolated DSL is expected to be stable.

Crucially, we have shown that there exists a symmetry-allowed coupling between lattice distortions at finite wavevectors $\bvec{K}_a/2$ (where $\bvec{K}_{1,2,3}$ are the corners of the hexagonal Brillouin zone) to spin-singlet, valley-triplet monopoles that act as order parameters for VBS ordering.
Considering static distortion fields, this coupling induces an instability of the DSL which can be understood as a direct analog of the spin-Peierls instability of the 1+1-dimensional Luttinger liquid, similarly driven by the proliferation of instanton events.

We have analyzed the coupling of monopoles to static lattice distortions first within conformal perturbation theory with a finite-temperature regularization, leading to a critical temperature $T_c \sim (g^2_c / K)^{1/(3-2\Delta_\Phi)}$ below which the DSL becomes unstable.
In a complementary approach, we have employed a scaling ansatz to find that the energy of  static lattice distortions $u$ lower the ground-state energy of the DSL $\Delta E \sim - u^\chi$ with $\chi < 2$, which generally outcompetes the harmonic elastic energy $E_\mathrm{elast} \sim u^2$, therefore favoring an equilibrium lattice distortion.

In both approaches, we found that the space of possible lattice distortion patterns and spontaneous VBS orderings exhibits accidental (continous) symmetries as artifacts of the low-energy theory.
Symmetry-allowed three-monopole operators may be dangerously irrelevant and can lift these accidental degeneracies, and we predict a resulting VBS order with a 12-site unit cell that preserves a $C_3$ symmetry of the lattice.

\begin{figure}
    \centering
    \includegraphics[width=.95\columnwidth]{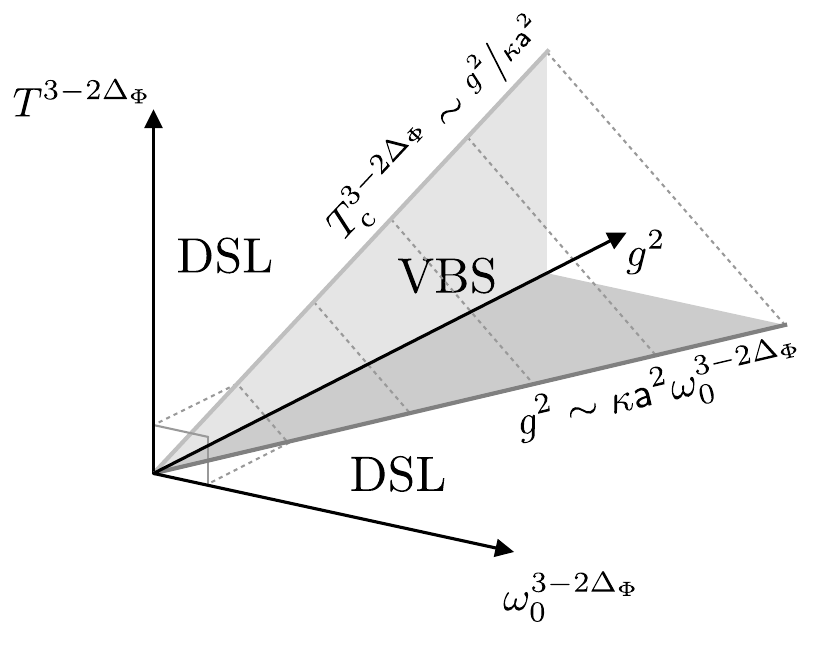}
    \caption{Scaling phase diagram for the Peierls-VBS instability of the Dirac spin liquid (DSL) on the triangular lattice, as a function of coupling $g$, temperature $T$ and frequency $\omega_0$. The shaded regions indicate the unstable parameter regimes, based on the finite-temperature calculation in the classical ($\omega_0 \to 0$) limit and the zero-temperature result for dynamical phonons ($\omega_0 > 0$). These two limits are seen to correspond, as the phonon frequency $\omega_0$ sets an effective temperature scale $T \sim \omega_0$ (as indicated by the dashed lines) which determines a correlation scale for critical fluctuations in the DSL.}
    \label{fig:pd_scaling}
\end{figure}

We have provided supporting evidence for our theoretical predictions via DMRG calculations for the $J_1$--$J_2$ model on the triangular lattice.
We manually distorted the lattice and modified its couplings depending on modified bond lengths from our predicted distortion and measure the ground state energy as a function of system size.
We found the in general quadratic energy to be strongest for 12-site pattern. Despite the difficulty of simulating strongly-frustrated triangular lattice Heisenberg models, we were able to observe signatures of ground state VBS order in agreement to the field-theoretical predictions.
Furthermore, we use finite-size scaling to deduce from \emph{within our microscopic model (UV scales)} that distortions at $\bvec{K}/2$ are strongly relevant perturbations to the DSL, as predicted by the low-energy IR field theory.

In addition to the conclusions drawn above, we highlight that this result provides compelling additional evidence of a gapless U(1) DSL ground state of the $J_1$--$J_2$ triangular lattice Heisenberg model and highlight the predictive power of the low-energy QED$_3$ theory for numerical simulations.
Despite computational limitations that introduce e.g., finite-size gaps, a detailed system-size scaling study can capture behavior characteristic of an underlying conformal fixed point. 
These methods have strong predictive power which is indispensable when looking to connect theory with experimental results.

Finally, we have considered the case of dynamic (non-dispersive) phonons, with some frequency $\omega_0 > 0$.
At finite frequencies, these degrees of freedom can be integrated out and yield an effective action containing retarded monopole-monopole interactions which have not been analyzed previously.
We showed that the spin-Peierls instability now occurs at finite couplings $g_c>0$, while in the stable regime the phonon spectrum is strongly renormalized.
We summarize our results in the `scaling phase diagram' shown in Fig.~\ref{fig:pd_scaling}, where we schematically indicate the parameter regime where an instability of the DSL is expected to occur.

\subsection{Application to other systems}
\subsubsection{Kagome lattice}

Our general framework and formalism can be straightforwardly applied to the $\Uone$ DSL state on the Kagome lattice \cite{ran07,song19}.
On the Kagome lattice, the UV symmetry quantum numbers of the monopoles are different than those on the triangular lattice.
In particular, the VBS monopoles carry lattice momenta $\bvec{M}_a$, corresponding to the Brillouin zone edge centers.
Further, under $C_6$ rotations, they pick up an additional phase: $s_a \Phi_a \rightarrow s_{a+1} \eu^{-2\pi\iu/3}\Phi_{a+1}$, and under reflection $R$ the monopoles transform as Table.~\ref{tab:UV-sym-mono} but with an additional Hermitian conjugation \cite{song19}.
The construction of the monopole-lattice coupling is complicated by the fact that the non-Bravais Kagome lattice has three sites per unit cells such that there are 6 (3 longitudinal + 3 transverse) distortion modes for a given lattice momentum.

\begin{figure}
\centering
\includegraphics[width=\columnwidth]{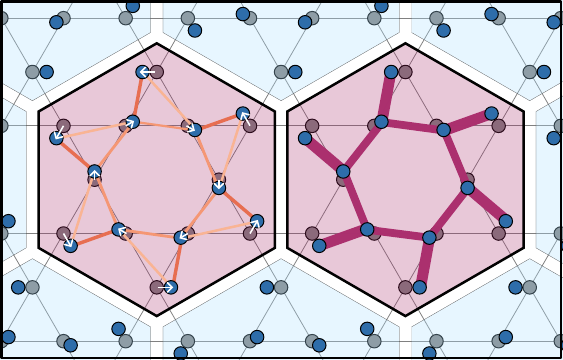}
\caption{Due to strong interactions of gapless gauge excitations in the U(1) DSL on the kagome lattice, it spontaneously distorts and forms 12-site pinwheel VBS pattern (coloured). Bond strengths and VBS order are shown as in Fig.~\ref{fig1}.}
\label{fig7}
\end{figure}

Based on the previous analysis, we can focus on the longitudinal modes on the sublattice $X=A,B,C$ (an intra-unit cell index) and momenta $\bvec{M}_a$ which we label as %
$u_X(\bvec{M}_a)$. Under the symmetry operations, the different (real) Fourier components with wavevectors $\bvec{M}_a$ are mapped onto each other (corresponding to changing monopole flavors).
Simultaneouly, the intra-unit cell coordinates  $X=A,B,C$ are appropriately mapped into each other (this intra-unitcell contribution is similar to the `pseudo-angular momentum' contribution of chiral phonons under $C_6$ as studied in Ref.~\cite{PhysRevB.100.094303}).
Hence, the $X$ components can be decomposed with respect to the irreducible representations (irreps) of $C_{3v} \simeq \mathcal{S}_3$ (the permutation group of three elements), which possesses three (complex) one-dimensional irreps which can be labeled by their $C_3$ eigenvalues.
Choosing the appropriate irrep, $h_a$, that cancels out the additional phase factors of the monopoles under $C_6$ rotations, allows us to write the couplings 
\begin{equation}
    S_g[\vec{h}] \sim  g\left(\eu^{2\pi\iu/3} h_1 \Phi_1 + 
    h_2 \Phi_2 -
    \eu^{-2\pi\iu/3} h_3 \Phi_3
    +\hc \right).
\end{equation} 
Having obtained the coupling, the results from Sec.~\ref{sec:instability} carry over straightforwardly; the coupling precipitates an instability towards VBS ordering and lattice distortion.
Transforming back to the sublattice basis $X=A,B,C$, the longitudinal phonon which couples to the monopoles is obtained by relative $2\pi/3$ phases $u_{X}(\bvec{M}_a) = \eu^{2\pi\iu\, X / 3} h_a$, allowing us to predict the lattice distortion as in Fig.~\ref{fig7}.
Identifying symmetry-equivalent expressions of the monopole operators in terms of microscopic spin bilinears as in \eqref{eq:vbsmonopolemapping}, we can additionally characterize the resulting ordered state on a microscopic level as the `pinwheel' VBS order \cite{PhysRevB.66.132408,Matan}.

\subsubsection{Square and honeycomb lattice}

On the square and honeycomb lattice, the $\Uone$ DSL is intrinsically unstable \cite{song19} due to the presence of relevant monopole operators that are singlets under all UV symmetries \footnote{Technically, this statement refers to the well-established staggered-flux and $0$-flux states on the square and honeycomb lattices that were studied in Ref.~\cite{song19}. As shown in Ref.~\cite{ye22}, there may exist stable DSL states on square and honeycomb lattice, though no parton constructions are known.}.
Furthermore, while stable $\Ztwo$ spin liquids exist on these lattices, they will not exhibit spin-Peierls instabilities due to the gapped nature of gauge excitations.
This applies equally to the gapless phase of the Kitaev honeycomb model, which is known to \emph{not} generically exhibit a spontaneous Kekul\'e distortion at weak coupling \cite{PhysRevB.99.155138,PhysRevB.101.245116}.

\subsubsection{Spin-orbital models: extending to general $N_f$}

Going beyond $S=1/2$, some spin-orbital models with intrinsic $\SUfour$ symmetry have been found to exhibit disordered ground states with algebraically decaying correlations \cite{corboz12,yamada18}.
Such spin-orbital liquids could be described by \QED3 with $N_f = 8$ fermion flavors \cite{calvera2021theory}.
Within the large-$N_f$ approximation, the monopole has dimension $\Delta_\Phi = 0.2651 N_f - 0.0381 + \mathcal{O}(1/N_f)$ \cite{dyer2013monopole,boro02}.
While this implies a lattice-monopole coupling is relevant for all $N_f<12$, we find it only precipitates an instability for $N_f <6$.
To see this, note that both the perturbative approach in Eq.~\eqref{eq:perturbative-energy} as well as the scaling ansatz [see Eq.~\eqref{eq:delta-e-scaling}] place a stronger constraint $\Delta_\Phi < 3/2$ for the system to undergo the spin-Peierls instability due to the harmonic energy cost of lattice distortions.
Hence we do not expect a weak-coupling spin-Peierls instability in the $\Uone$ Dirac spin-orbital liquid, despite the presence of relevant monopole operators.

\subsubsection{Generalization to $d$ dimensions}
Our results can be understood as 2+1-dim. extension of the spin-Peierls instability of interacting spin-half chains in 1+1 dimensions. 
We found an instability when lattice distortions couple to an operator with scaling dimension $\Delta_\Phi < d/2$, which is in accordance with the collapse of quantum criticality \cite{zach15,noad23} when $\nu (D+z) - 2 < 0$ in $D=d-1$ spatial dimensions with dynamical exponent $z=1$ and correlation length exponent $\nu^{-1} = d - \Delta_\Phi$ \footnote{Any linear coupling to monopoles, $\mathcal{S} \sim g h\, \Phi + \hc$, introduces a length scale to the otherwise scale-invariant theory, which we can identify as a correlation length $\xi$. By powercounting, we have $|g h| \sim \xi^{-(3-\Delta_\Phi)} \Leftrightarrow \xi \sim | g h|^{-1/(3-\Delta_\Phi)}$ from which we can read off the correlation length exponent $\nu$.}.
We emphasize that this result is more restrictive than the mere relevance of the monopole operator, $\Delta_\Phi < d$.

To illustrate this point, we can revisit the $(d=2+1)$-dim. CFT of free Dirac fermions, for concreteness as a low-energy theory of e.g., graphene or the honeycomb lattice Kitaev model.
While a Kekul\'e distortion $u$ couples via $g u \, \psi\overline\psi$ to the relevant fermion mass operator $M=\bar\psi \psi$ with $\Delta_M = 2$, the system is stable since $\Delta_M > 3/2$.
This stable free-fermion CFT is also obtained from \QED3 in the limit $N_f\rightarrow\infty$; we therefore stress that the spin-Peierls instability of the DSL crucially depends on strong gauge fluctuations in QED$_3$ at finite $N_f=4$.
This matches the intuition from interacting Luttinger liquids, where fermion interactions enhance the spin-Peierels instability compared to the free-fermion (equiv. to the spin-1/2 XY chain) due to instantons becoming more relevant.

\subsection{Outlook}
Our results for the DSL highlight that the coupling of quantum-critical states to other degrees of freedom can dramatically affect their stability.
In a similar vein, it will be interesting to study the impact of coupling other critical deconfined states to lattice degrees of freedom, such as deconfined quantum critical points \cite{senthil04,PhysRevB.97.195115,lee19} or non-Lagrangian QSL phases \cite{PhysRevX.11.031043}.
Such couplings may not only generate instabilities, but can also provide a route to resolve critical correlations via, e.g., anomalies in phonon spectra \cite{lee19}.

Turning towards experiments, we note that recently several triangular-lattice candidate materials have been identified, such as NaYbO$_2$ \cite{wilson19,tsirlin19} and YbZn$_2$GaO$_5$ \cite{xu2023realization}, which exhibit broad inelastic neutron-scattering spectra and, importantly, a $T^2$-scaling of the magnetic specific heat at low temperatures, suggestive of Dirac-type gapless excitations.

An open question on the road to confirmation of the realization in materials pertains to the fate of such state against quenched disorder \cite{PhysRevB.102.235165,PhysRevB.95.235146}, lattice defects, and additional degrees of freedom, such as interlayer coupling or indeed also phonons as studied here.
These are inevitably present in solid-state systems and a detailed understanding of their impact on QSLs will help in determining whether candidate materials display signatures of more trivial origin.

The VBS-ordered phase that is generated via the spin-Peierls instability is accompanied by a finite lattice distortion, which we expect to be measurable, e.g. by sensitive inelastic X-ray scattering experiments (as in spin-chain compounds \cite{PhysRevB.76.214304}).
The spin-Peierls mechanism could be disentangled from intrinsic structural distortions by studying the magnetic-field-dependence of critical temperatures \cite{hase93}.
It will be interesting to explore to what extent a spin-Peierls mechanism might be consistent with a recent report of a gapped ground state with finite structural distortion in an organic salt triangular lattice spin-liquid candidate material \cite{doi:10.1126/science.abc6363}.

Turning to the kagome lattice, pinwheel VBS ground states have been measured on the deformed kagome-lattice antiferromagnetic Heisenberg compound Rb$_2$Cu$_3$SnF$_{12}$ \cite{doi:10.1143/JPSJ.77.043707,Matan}. This material exhibits a lattice distortion which is compatible with the predicted spin-Peierls pattern of the DSL on the Kagome lattice. Although the presence of a lattice distortion at high temperature points towards a non-magnetic origin, it would be interesting to explore whether there are universal arguments that favor such pinwheel distortion patterns, possibly by considering incoherent high-temperature dynamics of the $\Uone$ DSL.
We note that although previous numerical studies of the kagome-lattice Heisenberg model with a simpler 12-site distortion pattern have suggested an unusually small nonzero critical coupling $\alpha_c>0$ \cite{PhysRevB.79.224417,PhysRevB.84.224411}, our zero-temperature results conclude that the pinwheel VBS is precipitated for any finite lattice-coupling $\alpha_c=0$ if the U(1) DSL is acting as an organizing or parent state for this material.

In real materials, phonons at finite momenta have non-zero frequency, and we have identified a power-law parameter regime where the spin-liquid is expected to be stable down to lowest temperatures.
Our study therefore places constraints on the realization and observability of the $\Uone$ Dirac spin liquid in materials.
Within stable DSL phase, a monopole-phonon coupling is responsible for a Kohn-type anomaly visible in the phonon spectrum. Therefore, phonon spectra may provide (indirect) insights into the critical correlations of a Dirac spin liquid, similar to elastic signatures of quantum critical points \cite{PhysRevLett.124.127205,PhysRevB.81.134438,noad23}.

It is in the purview of future numerical work to evaluate the stable parameter regime of the DSL and predict the magnitude of potential experimental signatures quantitatively, starting from realistic microscopic spin models and experimentally applicable phonon spectra.
The treatment of quantum phonons coupled to a QSL is especially computationally challenging but would allow for a more realistic modeling of candidate materials given exchange parameters and bare phonon frequency.

Our work also implies that the coupling to lattice degrees of freedom may provide a fruitful route to distinguish between different types of gapless spin liquids: Spin-singlet monopoles are reflected in observables such as phonon spectra (via the Kohn-like softening), or potentially Raman spectra (see also the recent proposals of Ref.~\cite{galitski23}).
In this context, we mention that the singular response of the DSL to \emph{externally} induced lattice displacements, in particular via strain, may provide a key signature for the experimental characterization of possible DSL phases.

The spin-Peierls instability of the Luttinger liquid in one-dimensional spin chains has been one of the paradigmatic examples of strong quantum fluctuations in low-dimensional quantum many-body systems.
The $\Uone$ DSL, described by a strongly interacting field theory, constitutes a similarly remarkable new phase of matter in two dimensions which we have shown also exhibits a spin-Peierls instability.
We hope that a better understanding of the stability of QSLs with respect to coupling to phonons or lattice disorder will help for the eventual discovery of these enigmatic quantum liquids in real materials. 

\acknowledgements

We thank L. Balents, F. Becca, A. Chubukov,  F. Ferrari, D. Hofmeier, Z.-X. Luo, J. Schmalian, S. Simon, R. Valent\'{i}, and A. Wietek for insightful discussions. 
We would like to especially thank F. Ferrari for discussions which helped us to identify the correct symmetric VBS pattern, as well as ongoing collaboration alongside F. Becca and R. Valent\'{i}.
U.F.P.S. and F.P. acknowledge support from the Deutsche Forschungsgemeinschaft (DFG, German Research Foundation), U.F.P.S. through a Walter Benjamin fellowship, Project ID~449890867 and F.P. through Germany’s Excellence Strategy EXC-2111-390814868. This research was supported by the National Science Foundation under Grant No. NSF PHY-1748958 and the European Research Council (ERC) under the European Union’s Horizon 2020 research and innovation program (Grant Agreement No. 771537). J.K. acknowledges support from the Imperial-TUM flagship partnership. The research is part of the Munich Quantum Valley, which is supported by the Bavarian state government with funds from the Hightech Agenda Bayern Plus. 

\textbf{Data and materials availability} – Code and data are available on Zenodo \cite{zenodoref}.

\appendix

\section{Luttinger liquid and XXZ model}\label{app:bozonization}
The XXZ model reads
\begin{equation}
    H_\mathrm{XXZ} = \sum_i J_{xy}(S^x_{i}S^x_{i+1}+S^y_{i}S^y_{i+1})
    + J_z S^z_{i}S^z_{i+1},
\end{equation}
where tuning $\Delta = J_z/J_{xy}$ tunes between the XY ($\Delta=0$) and AFM Heisenberg model ($\Delta=1$) \cite{giamarchi2004quantum}.
The bozonisation procedure maps this onto a continuum field theory 
\begin{multline}\label{eq:llfull}
  H[\phi(x)] = \frac{1}{2\pi} \int\dd[2]{x} \left[ K (\partial_t \phi)^2 + \frac{1}{K} (\partial_x \phi)^2 \right] \\
  + V_4 \int\dd[2]{x} \cos(4\phi),
\end{multline}
The kinetic term describes a free boson, where the Luttinger parameter $K$ characterizes the quasi-long range order and is also fixed by the microscopic bosonization mapping.
In the XY model, it takes the value $K=\frac{1}{2}$, and in the isotropic Heisenberg model it has the value $2$.

The term $V_4$ is an interaction strength of the instanton $\cos(4\phi)$.
An instanton tunnels between distinct ground states, introducing a winding in the periodic variable $\phi(x)$.
The symmetries of the original lattice constrain the possible interactions that can appear in the effective bosonized description.
Most relevant for our study, the discrete translational symmetry of the chain under $S_i\to S_{i+1}$ manifests as a constraint that the effective theory is invariant under $\phi\to\phi + \pi/2$. Therefore all interactions must be of the form of derivatives or periodic functions $V_n \cos(n\phi)$ with $n\ge4$.
Similarly, inversion symmetry requires the function be symmetric under $\phi(x)\to-\phi(x)$, which forbids terms of the form $W_n\sin(n\phi)$.

Given in terms of the microscopic spin Hamiltonian, the interaction is $\tilde{V}_4 = J^z/2$ at lattice length scales $\sim \mathsf{a}$.
The XY model with $J^z=0$ is therefore effectively described by a free massless boson at the UV scale.
To describe long-wavelength behavior, one must understand how couplings flow; to this end, we write $V_n(\zeta) = \zeta^{2-\Delta_n} \tilde{V}_n$ at length scales $\zeta \mathsf{a}$.
Expanding around the Gaussian fixed point, it can be shown that interactions of the form $\cos(n\phi)$ have a scaling dimension of $\Delta_n = n^2 \, K / 4$.
If an operator is relevant, then $\Delta_n<2$ and the coupling grows in the IR $\zeta\to\infty$.
Such an operator induces a mass scale in the spectrum $M(\zeta) = V_n(\zeta)^{\chi_n/2}=\zeta\,\tilde{V}_n^{\chi_n/2}$ with $\chi_n=2/(2-\Delta_n)$.

The instanton $\cos(4\phi)$ has dimension $\Delta_4 = 4K$, which for the Heisenberg model $K=1/2$ is marginally irrelevant
\cite{giamarchi2004quantum}. This means the system has a stable spin-liquid-like ground state with gapless excitations in the IR. 
The stability of this phase is protected by the lattice symmetries, as a more relevant instanton $\cos(2\phi)$ would be relevant at this Heisenberg point, but is forbidden on symmetry grounds.
For any lower $K<1/2$ (from $J_z>J_{xy})$, the interaction $V_4$ becomes relevant and the spectrum becomes gapped. In terms of the fermions, there is a Mott transition; in terms of the spins, the model enters the Ising universality class.

\section{Conformal field theory \& conformal data} \label{sec:CFT-intro}

In the following, we will refrain from working with the explicit formulation of the field theory $\mathcal{L}_{\mathrm{QED}_3}$ in terms of strongly interacting $\psi,a_\mu$ and having to account for monopoles as non-trivial background configurations.
Instead, we exploit the description of the IR fixed point as a conformal field theory (CFT) in 2+1 dimensions, which is characterized by its \emph{conformal data}, consisting of (1) a spectrum of scaling operators and (2) the operator product expansion.

\subsection{Scaling operators}
For $N=4$-flavor \QED3, the lowest-lying operators are given by the six charge-1 monopole operators.
The gauge-invariant combination of monopole operators and two Dirac zero modes $f_\alpha^\dagger$ reads \cite{song19}
\begin{equation} \label{eq:phi-ffm}
    \Phi^\dagger_{\alpha \beta} \sim f_\alpha^\dagger f_\beta^\dagger \mathcal{M}_{2 \pi}^\dagger,
\end{equation}
where $\alpha,\beta$ are $\SUfour$-indices. 
As written here, $\Phi_{\alpha\beta}^\dagger$ transforms in the antisymmetric rank-2 representation $\mathbf{6}$ of $\SUfour$.
It is convenient to use the isomorphism $\SO(6) = \SUfour/\Ztwo$, such that we can take $\Phi^\dagger_a$ with $a=1,\dots,6$ to transform as a six-dimensional vector.
A recent CFT bootstrap study \cite{alba22} estimates the scaling dimension $\Delta_\Phi \in (1.02,1.04)$, very similar to the large-$N$ (subleading order) result $\Delta_\Phi \approx 1.02$ \cite{chester16}.
Other numerical works give similar estimates for $\Delta_\Phi$ \cite{karthik16,he22}.

The next lowest-lying operators are the adjoint fermion masses
\begin{equation}
    M^{\alpha \beta} = \overline{\psi} \sigma^\alpha \tau^\beta \psi
\end{equation}
with $\alpha,\beta =0,1,2,3$ and $\sigma$/$\tau$ act on spin/valley indices of the fermions. 
The masses $M^{\alpha \beta}$ transform in a sixteen-dimensional reducible representation of $\SUfour$, which splits into the singlet and adjoint irreducible representations $\mathbf{16}=\mathbf{1}\oplus\mathbf{15}$.

The scaling dimension of the adjoint mass term $M^{\mu\nu}$ (with $\mu = \nu=0$ excluded) is strongly relevant, with $\Delta_M = 1.46$ in the large-$N_f$ expansion \cite{chester16} and $\Delta_M\in(1.33,1.66)$ according to a recent bootstrap study \cite{alba22}.
Due to their non-trivial representations under lattice transformations, these perturbations cannot trivially be added to the action. 
On the triangular lattice the symmetry-allowed four-fermion interaction is believed to be irrelevant \cite{Di_Pietro_2017}.

The singlet $M_{00}$ corresponds to a chiral mass term and can be tuned to produce a chiral spin liquid state. In large-$N_f$, this term is irrelevant $\Delta_0 = 3.08$, and thus it is believed that spontaneous chiral symmetry breaking does not occur \cite{apple88,grover14,janssen14}.

\subsection{Operator product expansion}
The operator product expansion (OPE) states how operators approaching each other are expanded in the basis of primaries as
\begin{equation} \label{eq:def-ope}
    \lim_{x \to y} \mathcal{O}_i(x) \mathcal{O}_j(y) = \lim_{x\to y} \sum_k \frac{C_{ij}^k}{|x-y|^{\Delta_i + \Delta_j - \Delta_k}} \mathcal{O}_k(y),
\end{equation}
where the $C_{ij}^k$ are the OPE coefficients.
If the CFT is endowed with some global symmetry group $\mathcal{G}$, each term on the RHS of \eqref{eq:def-ope} will transform in some irreducible representation of $\mathcal{G} \times \mathcal{G}$.
We stress that Eq.~\eqref{eq:def-ope} corresponds to the asymptotic formulation of the OPE, but in general, OPEs can be shown to be \emph{convergent} series if all primaries and their descendants are included.
The full asymptotic form of the operators can be written in terms of the two real OPE coefficents $C_{\Phi\Phi}^M$ and $C_{M\Phi}^\Phi$ in the following way:
\begin{subequations}
\begin{align}
&\Phi_a^\dagger(x) \Phi_b(y) \sim \frac{\delta_{ab}}{|x-y|^{2 \Delta_\Phi}} + \frac{\iu C_{\Phi\Phi}^M }{|x-y|^{2 \Delta_\Phi- \Delta_M}}\mathcal{F}^{ab}_{\mu \nu} M^{\mu \nu}(y) + \dots  \\
&\Phi_a(x) M^{\mu \nu}(y) \sim \frac{\iu C_{\Phi M}^\Phi }{|x-y|^{\Delta_M}} \bar{\mathcal{F}}_{ab}^{\mu\nu}\Phi_b(y) + \dots,
\end{align}
\end{subequations}
where the tensor $\mathcal{F}^{ab}_{\mu \nu}$ maps elements of the adjoint-$\SUfour$ representation to rank-2 antisymmetric representations of $\SO(6)$ and $\bar{\mathcal{F}}_{ab}^{\mu\nu}$ its inverse. Implicit sums over $(\mu,\nu)$ are taken to exclude the $(0,0)$ element. The nonzero components are $\mathcal{F}_{0i}^{ab} = \epsilon^{abi}$ for $a,b\le3$; $\mathcal{F}_{0i}^{ab} = \epsilon^{a-3,b-3,i}$ for $a,b>3$; and $\mathcal{F}_{ij}^{ab} = \delta_j^a\delta^b_{i+3}-\delta_{i+3}^a\delta^b_{j}$, as defined in Ref.~\cite{luo22}. Here, the components $i,j=1,2,3$. We can also write the components of its inverse $\bar{\mathcal{F}}_{ab}^{\mu\nu}=\mathcal{F}^{ab}_{\mu \nu}/2$.

\section{Fermion masses coupled to the lattice}\label{app:massterms}

{\setlength{\tabcolsep}{10pt}\renewcommand{\arraystretch}{1.2}\begin{table}
\caption{\label{tab:UV-sym-mass}Discrete symmetry transformations of fermion mass operators.}
\begin{tabular}{lccccc}
\hline\hline
    & $(T_1,T_2)$ & $R$ & $C_6$ & $\mathcal{T}$ 
     \\ \hline
\noalign{\vskip 0.5mm}
     $M_{00}$ & $(+,+)$ & $-$ & $+$ & $-$ \\ 
\noalign{\vskip 1mm}
$M_{i0}$ & $(+,+)$ & $+$ & $-$ & $+$ \\ 
\noalign{\vskip 1mm} 
$M_{01}$ & $(-,-)$ & $+M_{03}$ & $-M_{02}$ & $+$ \\ 
     $M_{02}$ & $(+,-)$ & $-M_{02}$ & $+M_{03}$ & $+$ \\ 
     $M_{03}$ & $(-,+)$ & $+M_{01}$ & $+M_{01}$ & $+$ \\ 
     \noalign{\vskip 1mm}
     $M_{i1}$ & $(-,-)$ & $-M_{i3}$ & $+M_{i2}$ & $-$ \\ 
     $M_{i2}$ & $(+,-)$ & $+M_{i2}$ & $-M_{i3}$ & $-$ \\ 
     $M_{i3}$ & $(-,+)$ & $-M_{i1}$ & $-M_{i1}$ & $-$ \\
     \noalign{\vskip 0.5mm}
     \hline\hline
\end{tabular}
\end{table}}

\subsection{Effective action including fermion mass couplings}
In the main text, we constructed an effective action of VBS monopoles coupled to lattice distortions by considering the leading contributions from a redefinition of the lattice $\bvec{R}_i \to \bvec{R}_i + \bvec{u}_i$.

In this section we will approach the task of constructing an effective theory of \QED3 coupled to a lattice distortion using a `top-down', symmetry-based approach. This path allows us to categorize couplings to all possible operators in the theory which could contribute to a ground state instability. This approach validates the previous method by showing that the monopole indeed causes the leading instability and that the subleading operators couple to the same momentum modes of the lattice distortion.

We now write down a general set of couplings to the leading relevant operators in \QED3: the monopoles $\Phi_a$ ($a=1\dots6$) and adjoint fermion masses $M_{\mu\nu}$ [$\mu,\nu=0\dots3$, $(\mu,\nu)\neq(0,0)]$. The general form of this action is
\begin{equation}\label{eq:acwithmasses}
    S \sim \int\dd[3]{x} \left[ \sum_{a=1}^6 \left(\varphi_a^* \Phi_a(x) +\mathrm{h.c.}\right)+\sum_{\mu,\nu} m_{\mu\nu} M^{\mu\nu}(x)  \right].
\end{equation}
The monopoles transform under the UV symmetries as listed in Table~\ref{tab:UV-sym-mono}, and the masses as listed in Table~\ref{tab:UV-sym-mass}.
We will preserve the lattice symmetries by constructing an appropriate object $h_a$ that transforms in the same representation as $\Phi_a$ of the IR emergent symmetry group $\SUfour$. Because the UV lattice symmetries embed in this emergent group, constructing an invariant $\SUfour$-scalar is equivalent to writing an function that is invariant under lattice symmetries.

For our physical system of interest, the question becomes whether it is possible to construct a vector $h_a$ that transforms as the monopole under physical symmetries in Table~\ref{tab:UV-sym-mono} out of the physical lattice-distortion vector field $\bvec{u}(\bvec{x})$. 
This object will naturally transform as a vector under the $\SOthree_{\mathrm{valley}}$ subgroup of the full $\SO(6)$, and not transform under $\SOthree_{\mathrm{spin}}$. 

To classify all perturbations that could be symmetry-allowed in the presence of a deformation vector $\bvec{u}_{\bvec{q}}$, we must restrict our search by using the following criteria: 
(1) consider distortions which are periodic with wavevector $\bvec{q}$ that is equal to that of the field theory operator and (2) are fully invariant under the discrete subgroups $C_6$, $\mathcal{T}$. These two requirements will be equivalent to restricting to coupling field theory operators to fields which are vectors under $\SOthree_{\mathrm{valley}}$ and trivial under $\SOthree_{\mathrm{valley}}$.
Finally, by way of the correspondence between vector and adjoint representations, one can write down an object bilinear in the lattice deformation field which can couple to field theory objects in the $\SOthree_{\mathrm{valley}}$-adjoint representation.

We satisfy the above constraints by defining field $h_a$ as
\begin{equation}
    h_a = \iu s_a (\bvec{k}_a \cdot \bvec{u}_{\bvec{k}_a}),
\end{equation}
where we have chosen $\bvec{k}_a$ as a basis of three vectors separated by $2\pi/3$ angles. 
It satisfies translational invariance since it has the same momentum eigenvalue as the monopole, and also transforms as a vector under $\SOthree_{\mathrm{valley}}$.
Under IR symmetries it transforms as the VBS monopole, as listed in Table~\ref{tab:UV-sym-mono}.
We see it is possible to preserve the full lattice symmetry group by coupling this field to the monopoles and masses as follows:
\begin{equation}\label{eq:mcplform}
    \varphi_a^* = g \, h_a^*, \quad
    m_{0a} = \iu m \, \mathcal{F}^{bc}_{0a} h_b^* h_c,
    \quad a=1,2,3,
\end{equation}
where the non-zero components give $m_{0a} = \iu m \, \epsilon^{abc}h_b^* h_c$.

We can exclude any direct linear coupling of lattice distortions to spin-triplet monopoles $\Phi_a$, $a=4,5,6$. In terms of the discrete symmetries this follows since under the $C_6$ transformation $\Phi_a\to-\Phi_a^\dagger$. This is incompatible with any coupling to a vector $\bvec{u}_a$, which necessarily rotates under such an operation $\bvec{u}_{a}\to-\bvec{u}^*_{a+1}$.

We could instead attempt to couple distortions $h_a$ directly to the valley-triplet fermion masses $M_{ia}$, and then generate monopole couplings by using the inverse mapping between 2-dimensional and adjoint representations. However these masses also transform in the index $i$ as a spin-vector, which could provide a hindrance to writing an invariant action.
Inspecting the behavior under lattice symmetries (Table~\ref{tab:UV-sym-mass}), we find the coupling satisfies translational invariance (1) when coupling to the (real) modes at the $\bvec{M}_{1,2,3}$-points on the Brillouin zone edges $\tilde{\bvec{u}}_a = {\bvec{u}}_{\bvec{M}_{a}}$. We are therefore motivated to write the couplings $\tilde{m}_i,\tilde{g}_i$ for $i=1,2,3$,
\begin{equation}
    m_{ia} = \tilde{m}_i \tilde{h}_a, \quad 
    \varphi_a = -\iu \tilde{g}_i\,  {\mathcal{F}}_{bc}^{ia}
    \tilde{h}^b\tilde{h}^c,
    \quad a=1,2,3,
\end{equation}
with non-zero components giving $\tilde\varphi_{a+3} = -\iu \tilde{g}_b \tilde{h}_a \tilde{h}_b$.
However there is no choice of $\tilde{h}_a$ which leaves $m_{ia} M^{ia}$ invariant under both discrete lattice symmetries (2). 
For this reason, we must exclude a direct coupling between fermion masses and the $\bvec{M}$ points from the effective action.

We are now able to write the full symmetry-allowed action at lowest order in relevant operators as a sum of VBS monopole and mass terms,
\begin{equation}\label{eq:acwithmasses_vbsonly}
    S_I[h_a] =\int\dd[3]{x} \left[ g \left(h_a^* \, \Phi_a(x) +\mathrm{h.c.}\right)
    +
    \iu m \, \epsilon^{abc} h^*_b h_c \, M_{0a}(x) \right],
\end{equation}
both of which are coupled to lattice distortions via $h_a = \iu s_a (\bvec{k}_a \cdot \bvec{u}_{\bvec{k}_a})$.
There is furthermore a set of compound operators which may be added as deformations to the theory. One example is the triple-monopole term $\Phi_1\Phi_2\Phi_3$; allowed even on the undistorted lattice, this term is irrelevant at the conformal IR fixed point.
One can equally construct double-monopole terms coupled to a commensurate lattice distortion (or appropriate combination thereof).
At this point we will ignore such compound operators which have a scaling dimension $\Delta>3/2$; this is justified by seeing that the leading IR divergence contribution to the free energy from such an operator goes as $\beta^{3-2\Delta}$ (at quadratic order in perturbation theory).

\subsection{Perturbative action}
We will now derive the subleading contribution to the perturbative free energy functional due to lattice-fermion mass couplings. Performing the same calculation as in the main text for lattice-monopole couplings, we enter the weak coupling regime $g^2 \beta^{3-2\Delta_\Phi} |\vec{h}| \ll 1$ and perform a perturbative expansion of the free energy $\langle S_m^2\rangle /\beta V$. Using the explicitly derived structure of the coupling, we find
\begin{align}
    \frac{\langle S^2_m \rangle_{\mathrm{QED}_3}}{\beta V} &= -m^2\int\dd[3]{x} \,  \frac{\epsilon^{abc}\epsilon^{ade} \, h_bh_{d}^* \, h_ch_{e}^*}{|x|^{2\Delta_M} } \nonumber\\
    &= c_{\Delta_M} m^2 \beta^{3-2\Delta_M} \left( |\vec{h}\cdot\vec{h}|^2 - |\vec{h}|^4 \right).
\end{align}
This is IR-divergent if the exponent $\Delta_M<3/2$, which is compatible with large-$N$ calculations and within the range of values suggested by the conformal bootstrap \cite{alba22}.
Assuming this condition is satisfied, this subleading divergent contribition will be minimised by states satisfying $\vec{h}\cdot\vec{h} = 0$.

If this condition $\Delta_M <3/2$ is not satisfied, we would still expect the perturbative action to have only a $\SOthree$ symmetry which is inherited from the $\SOthree_{\mathrm{valley}}$ symmetry of the effective action. This inevitably occurs at higher order in perturbation theory, where the monopole-monopole and monopole-mass terms produce additional divergent contributions proportional to $\mathcal{C}(\beta)(|\vec{h}\cdot\vec{h}|^2 - |\vec{h}|^4)$ \cite{luo22}. 
At quadratic order there is a contribution from the $M\times M$ OPE channel above, giving $\mathcal{C}^{(2)}(\beta) = c_2\, \beta^{3-2\Delta_M}$.
At higher order there are contributions from the following channels: $\Phi^\dagger\times (M\times \Phi) \to \Phi^\dagger\times\Phi$ is asymptotically 
$\mathcal{C}^{(3)}(\beta) = c_3 \,\beta^{6-2\Delta_\Phi-\Delta_M}$; 
and the channel $(\Phi^\dagger\times\Phi)\times (\Phi^\dagger\times\Phi) \to M\times M$
contributes
$\mathcal{C}^{(4)}(\beta) = c_4\, \beta^{9-4\Delta_\Phi}$.
We can constrain $c_2\sim m^2>0$ and $c_4\sim g^2 (C_{\Phi\Phi}^M)^2 > 0$ to be positive, but the sign of $c_3$ cannot be fixed in this way.
There will be multiple additional contributions at this order (e.g. from $\langle S_g^2\rangle^2/\beta V$) which produce the potential $|\vec{h}|^4$, leaving the sign of this coefficient in the effective action indeterminate. However we predict that the leading contributions to the $|\vec{h}\cdot \vec{h}|^2$ term is positive.

\section{Numerical Simulations}\label{app:convergence}
\label{appendix:numerical_simulations}

\begin{figure}
    \centering
    \vspace{2mm}
    \includegraphics{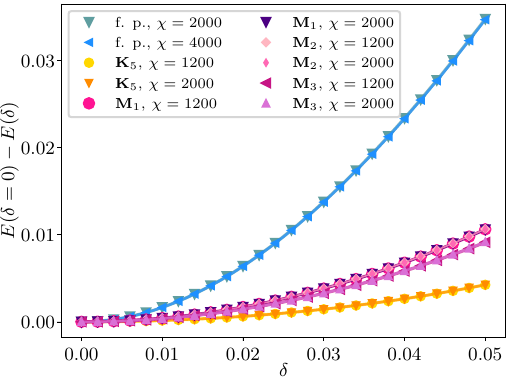}
    \caption{Convergence of the DMRG simulations for different patterns as shown in Fig.~\ref{fig:energygain} (a) on an $L=6$-cylinder. We show the energy gain for the full pattern (f. p.) as defined in Eq.~\eqref{eq:rs12site}, for the three-sublattice pattern for ${\bvec K}_5 = - {\bvec K}_3$ with a phase of $\theta = \frac{1}{2}$ and for patterns associated to the midpoints ${\bvec M}_i$ ($i=1,2,3$) of the edges of the Brillouin zone with phases $\theta_i = \frac{1}{6}, \frac{1}{2}, \frac{5}{6}$ respectively. Fig.~\ref{fig:energygain} shows the average over the different orientations of the $\bvec M$-point patterns. Calculations on the $L=6$ cylinder have been performed by the flux-insertion method and a bond dimension of up to $\chi=4000$; convergence is evident from the independence of the response on bond dimension. The distorted couplings in real space for the ${\bvec K}$-pattern is shown in Fig.~\ref{fig:distortion_patterns_energy_response} (a), for the ${\bvec M}$-points in Fig.~\ref{fig:Numerical_appendix_patterns} (c) and Figs.~\ref{fig:distortion_patterns_energy_response} (b--c).}
    \label{fig:convergence_full_pattern_all_M}
\end{figure}

We use the density matrix renormalization group algorithm (DMRG) \cite{White1992, Schollwoeck2011, Hauschild2018} on infinite cylinders \cite{McCulloch2008, Stoudenmire2012, Gohlke2017} to obtain the ground state of the $J_1$--$J_2$ Heisenberg model on a triangular lattice. The lattice is wrapped onto the cylindrical geometry by closing the boundary conditions periodically along the circumference $L_y\equiv L$. There are several possible ways that the periodic closing of the lattice can be achieved. In this work, we use the $YC{L_y}-0$ geometry \cite{Zhu2015, Hu2019}, where $n=0$ determines the detailed boundary condition. The lattice sites ${\bvec r}$ and ${\bvec r} + L_y {\bvec a}_2 - n {\bvec a}_1$ are thereby identified. We use $U(1)$-charge conservation for all numerical simulations discussed in this work. We use the setup of infinite matrix product states (MPS) \cite{Schollwoeck2011, Hauschild2018} and optimize the ground state running infinite DMRG (iDMRG) on unit cells of size $L_y \times 3$, where $L_x = 3$ denotes the number of rings in the cylinder geometry.

\begin{figure}
\centering
\includegraphics{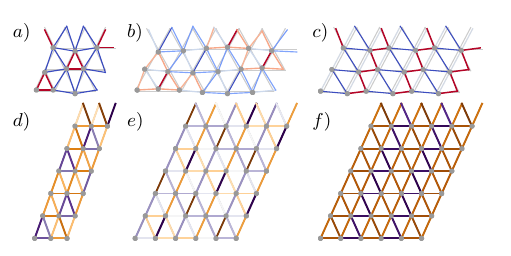}
\caption{Top subfigures (a--c) show the modified nearest-neighbour Heisenberg exchanges under the lattice distortion patterns ${\bvec K}_3$, $-{\bvec k}_3 \equiv \bar {\bvec k}_3$ and ${\bvec M}_3$ with the phases $\theta\left({\bvec K}_3\right) = \frac{1}{6}$, $\theta\left(\bar{\bvec k}_3\right) = \frac{1}{6}$ and $\theta\left({\bvec M}_3\right) = \frac{5}{6}$. Bottom subfigures (d--f) show the respective nearest-neighbor ground state correlations $\langle\vec{S}_i\cdot\vec{S}_j\rangle$ for a distortion of $\delta = 0.0038$, with purple showing more negative VBS weight. For each subfigure (d--f), the color scale has been normalized to the minimal and maximal value of the correlations measured. The patterns shown here have been used to study the finite size effects as discussed in Fig.~\ref{fig:scaling_analysis}.}
\label{fig:Numerical_appendix_patterns}
\end{figure}

\begin{figure}
    \centering
    \includegraphics{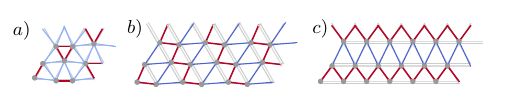}
    \caption{Additional patterns as used for the DMRG simulations generated from the following momenta: (a) $-{\bvec K}_3$ with phase $\theta = \frac{1}{2}$, (b) ${\bvec M}_1$ with phase $\theta = \frac{1}{6}$ and (c) ${\bvec M}_2$ with $\theta = \frac{1}{2}$. The corresponding energy responses under increasing $\delta$ are shown in Fig.~\ref{fig:energygain} and Fig.~\ref{fig:convergence_full_pattern_all_M}.}
    \label{fig:distortion_patterns_energy_response}
\end{figure}

When optimizing the ground state of the two-dimensional Hamiltonian on the lattice, special care has to be taken when we consider a geometry with an even circumference $L_y$. The isotropic $J_1-J_2$ Heisenberg model on even cylinders comprises two different topological sectors in the putative quantum spin liquid phase \cite{Hu2015, Zhu2015}. The iDMRG algorithm in this case finds the even sector in which the entanglement spectrum on the bonds is symmetric around the total $S_z$-quantum number $q_z=0$. One can transition into the odd sector by adiabatically inserting a flux of $2\pi$ through the cylinder \cite{Hu2019}. In this process, the couplings of the model acquire complex phases along the circumference. The final state's entanglement spectrum is symmetric around $q_z = \frac{1}{2}$ which can be understood as a spinon quasiparticle with $S_z=\pm \frac{1}{2}$ residing on each boundary \cite{Hu2015}. To find the ground state for the $L_y=6$-cylinder, we apply this protocol referred to as the flux-insertion method.

Alternatively, one can enforce a preference for odd dimer coverings on cylinders with even $L_y$ by omitting a single site on each of the outermost cylinder rings of the unit cell for the first iDMRG sweeps before restoring the full model \cite{Zhu2015, Hu2019} and having the algorithms optimize the ground state MPS until convergence. We refer to this protocol in contrast to the flux-insertion as the odd-sector method.
We ensured that for sufficiently large bond dimensions, the energies obtained on $L_y=6$ by the two distinct methods described above agree.

Starting from the ground state of the undistorted model, we can adiabatically turn on the distortion parameter $\delta$ and modify the couplings of the Hamiltonian according to the pattern generated by a certain momentum ${\bvec Q}$. The parameter $\delta$ controls the strength of the distortion.
The full 12-site unit cell pattern in real-space is obtained from equation~\eqref{eq:rs12site} as follows:
\begin{equation}
    {\bvec r}_i = {\bvec R}_i + \frac{2\delta}{\sqrt{3}} \sum_{a=1,2,3} s_a \hat{\bvec{k}}_a \, 
    \sin\left[\bvec{k}_a\cdot \bvec{r}_i + \frac{2\pi\, (a-1) }{3}\right],
\end{equation}
where ${\bvec R}_i$ denotes the undistorted position of site $i$ and $\delta$ measures the magnitude of the distortion.
All other patterns are generated via the formula
\begin{equation}
    {\bvec r}_i = {\bvec R}_i + \delta
    \begin{pmatrix}
        \cos[{\bvec Q} \cdot {\bvec R}_i + \theta]\\
        \sin[{\bvec Q} \cdot {\bvec R}_i + \theta]
    \end{pmatrix}
\end{equation}
where the phase $\theta$ is an additional parameter that gives rise to the various patterns for each momentum ${\bvec Q}$ discussed in Section~\ref{sec:energy_response_to_distortion}. The precise phases used for different distortion patterns are given in Figs.~\ref{fig:Numerical_appendix_patterns}~and~\ref{fig:distortion_patterns_energy_response} alongside the real-space modification of the couplings (for a value $\delta = 0.1$ for demonstration purposes).

For the investigation of the system size dependence, the simulated pattern at the commensurate monopole distortion for a single $\frac{{\bvec K}}{2}$-momentum was taken as the negative $-{\bvec k}_3 = \frac{{\bvec K}_3}{2}$ with a phase of $\theta = \frac{1}{6}$. For the sake of readability, the corresponding data points in the figures have been named ${\bvec k}_3$. The same naming convention has been chosen for the discussions in the main text. The pattern generated by ${\bvec k}_3$ in contrast to $-{\bvec k}_3$ would be mirrored along the ${\bvec a}_2$-axis of the lattice, which is an irrelevant modification in terms of the physics of the system.

Fig.~\ref{fig:energygain} shows the energy response for various patterns under adiabatic increase of the distortion $\delta$. The ${\bvec M}$-point patterns as shown in Fig.~\ref{fig:Numerical_appendix_patterns}(c),~ Figs.~\ref{fig:distortion_patterns_energy_response} (b--c) have different orientations on the cylinder with slightly differing energy responses. Fig.~\ref{fig:energygain} therefore shows the averaged energy response over all three orientations.
The shaded area indicates the range of the energy gain for the different orientations on the cylinder as shown explicitly in Fig.~\ref{fig:convergence_full_pattern_all_M}.

\begin{figure}
\centering
\vspace{2mm}
\includegraphics{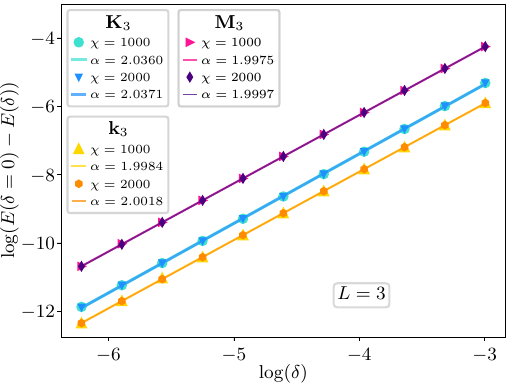}
\caption{Convergence of the DMRG simulations for $L=3$ and the three patterns generated by the momenta ${\bvec K}_3$, $-{\bvec k}_3$ and ${\bvec M}_3$ for various bond dimensions. The real-space patterns are shown in Fig.~\ref{fig:Numerical_appendix_patterns}.}
\label{fig:Numerical_appendix_L3}
\end{figure}

\begin{figure}
\centering
\includegraphics{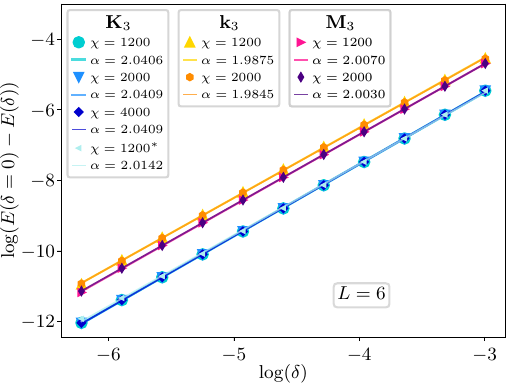}
\caption{Convergence of the DMRG simulations for $L=6$ and the three patterns generated by the momenta ${\bvec K}_3$, $-{\bvec k}_3$ and ${\bvec M}_3$ for various bond dimensions. For the pattern generated by ${\bvec K}_3$, the ground state has been obtained by adiabatically inserting a flux of $2\pi$ through the cylinder except for the curve marked with an asterisk. For this simulation and all other patterns shown, the odd sector method was used (cf. \ref{appendix:numerical_simulations}).}
\label{fig:Numerical_appendix_L6}
\end{figure}

To check convergence in the virtual bond dimension $\chi$ of the matrix product state, we can store the state along with the environments of the infinite DMRG simulation and restart the optimization under an increase of the bond dimension. The saving of the environments is necessary to ensure that the algorithm does not fall back into a possibly wrong topological sector.

In Fig.~\ref{fig:convergence_full_pattern_all_M} and Figs.~\ref{fig:Numerical_appendix_L3}--\ref{fig:Numerical_appendix_L6}, we show the convergence for $L=3$ and $L=6$ for various patterns. The real-space distortions of the couplings alongside the spin-spin correlations $\langle {\vec S}_i \cdot {\vec S}_j \rangle$ resulting from the numerically obtained ground states are shown in Fig.~\ref{fig:Numerical_appendix_patterns}. As can be deduced from the figures, the energy difference is already extremely well-converged for intermediate bond dimensions of a few thousand. The absolute ground state energy, however, is not necessarily so well-converged for all circumferences and bond dimensions. Despite this, we have confidence in our results for the energy gain (even for small values $\mathcal{O}(10^{-10})$) because of the stricter convergence criteria applied to the DMRG algorithm.

Note that the size of the unit cell in $x$-direction, denoted by $L_x$, needs to be chosen in accordance with the pattern we simulate. More precisely, whereas the patterns for ${\bvec K}$-points can be fitted on any multiple of three, i.e. $L_y\times 3$ is sufficient for the unit cell size to have a commensurate geometry (given that $L_y$ itself is a multiple of three), the distortion patterns generated by the ${\bvec M}$-points or $\frac{\bvec K}{2}$ momenta, require an extent of $L_x=6$ (or multiples thereof).

\begin{figure}
\centering
\includegraphics{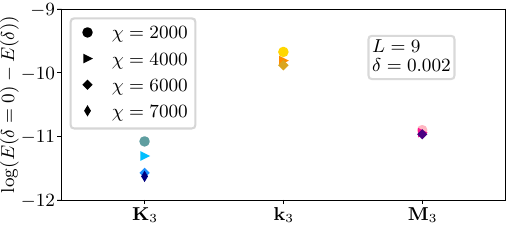}
\caption{Convergence of the DMRG simulations for $L=9$ for a specific small distortion of $\delta = 0.002$ and the three patterns generated by the momenta ${\bvec K}_3$, $-{\bvec k}_3$ and ${\bvec M}_3$ for various bond dimensions. The phases are chosen as defined in Fig.~\ref{fig:Numerical_appendix_patterns}.  The circles denote data points for bond dimension $\chi = 2000$, the triangles for $\chi = 4000$, the diamonds for $\chi = 6000$ and the thin diamonds for $\chi=7000$. Note the logarithmic scale on the y-axis.}
\label{fig:Numerical_appendix_L9}
\end{figure}

In contrast to $L_y = 3$ and $L_y=6$, the circumference of $L_y=9$ is more challenging as the computational complexity grows exponentially in $L_y$. In particular, the DMRG simulation tends to converge to semistable states that can change abruptly when bond dimension is insufficient. For small distortion strengths $\delta$, the convergence is more stable for the accessible bond dimensions.
In Fig.~\ref{fig:Numerical_appendix_L9}, we show data for various patterns for $L_y=9$, which appear to converge for two of the three patterns studied for a bond dimension of $\chi = 6000$ and in the case of the ${\bvec K}$-point for $\chi = 7000$.

\section{Phonon spectral function}\label{sec:phononspecfn}
In this section, we will evaluate the effect of fluctuations of the stable, quantum-critical DSL phase on the spectrum of the phonons. At zero-temperature we find a continuous spectrum of phonons \cite{lee19} with a divergent spectral weight as $\omega\to0$. 
As large temperature, the analysis in the main text shows there should instead be a well-defined pole in the phonon propagator with an energy that corresponds to a renormalized phonon frequency.
We expect this pole to move towards zero as the spin-Peierls temperature is approached.
We supplement the perturbative self-consistent analysis of the phonon self energy [given in Eq.~\eqref{eq:g-renorm}] with a full evaluation of the phonon spectral function. This provides a confirmation that the spectrum is described by a Kohn-anomaly-like dip at large temperatures, which will go to zero frequency at the spin-Peierls transition.

\begin{figure}[h]
\centering
\includegraphics[width=\columnwidth]{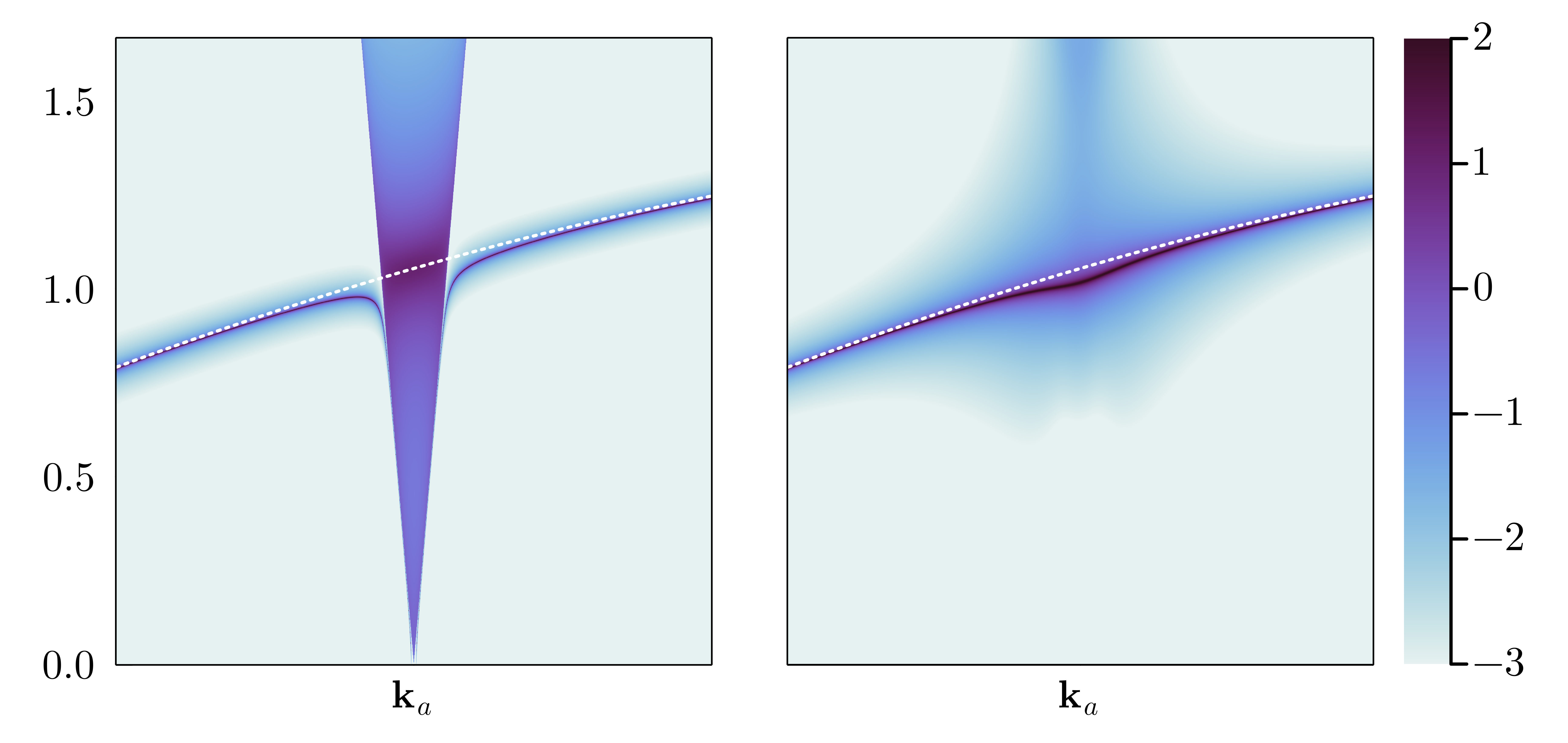}
\caption{Phonon spectral function $S(\bvec{k},\omega)$ plotted for zero temperature (left) and $T=3\,\omega_0$ (right) along the momentum slice between $\Gamma$ and $\bvec{K}_a$ and for energies $\omega/K$. Plots produced with $\gamma=0.3$, $g=0.3$, and $c=15$. White dotted line shows the un-renormalized phonon dispersion $\omega_0(\bvec{k})$; color axis is logarithmic.}
\label{spectra}
\end{figure}

To describe this effect, we first effectively integrate-out the VBS-monopole fluctuations of the DSL to produce the dressed propagator of the phonons at quadratic order \cite{lee19},
\begin{equation}\label{eq:phononcorrec}
    G(\omega,\bvec{k}) = \frac{1}{\omega_0(\bvec{k})^2 - g^2 |\bvec{k}_a|^2\rho^{-1} \chi(\omega,\bvec{k}) -\omega^2 }
\end{equation}
where the bare phonon dispersion $\omega_0(\bvec{k})^2 = \mathcal{K}_{\bvec{k}}^2/\rho$ [Eq.~\eqref{eq:distcost}]. 
The phonon spectral function is given by $S(\omega,\bvec{k}) = 2 \Im D(\omega+ \iu\epsilon, \bvec{k})$. 
The spin-VBS channel susceptibility has the following form at zero temperature $\chi(\omega,\bvec{k}) = [c^2|\bvec{k}_a- \bvec{k}|^2-\omega^2]^{-(3/2 - \Delta_\Phi)}$ featuring a divergent continuum of excitations at $\omega>  c |\bvec{k}-\bvec{k}_a|$ (where $c$ is some emergent ``speed of light''), but is not generally known at finite temperature. One can write generally $\chi(\omega,\bvec{k}) = T^{2\Delta_\Phi-3}\, \Phi({\omega}/{T},{c|\bvec{k} - \bvec{k}_a|}/{T}),$
where $\Phi$ is an unknown universal scaling function and constrain it in the large-temperature limit \cite{PhysRevLett.69.2411,sachdev_2011,PhysRevLett.114.177201}.
While even asymptotic results for the frequency-dependent susceptibility at finite-$T$ are not exactly known, for illustrative purposes we heuristically assume a form similar to the expansion obtained by Sachdev and Ye \cite{PhysRevLett.69.2411}.
In Figure.~\ref{spectra}, we plot the phonon spectral function, in the exact, zero-temperature limit (left) as well as a high-$T$ asymptotic result (right). We expect a temperature-dependent softening of the phonon mode towards zero at the spin-Peierls temperature. This is accompanied by the emergence of a sharp divergent continuum of excitations which blurs with the mode at low energies. A better understanding of the scaling form $\Phi$ would allow a prediction of how this mode approaches zero, a feature which can be experimentally and numerically probed.

\bibliographystyle{apsrev4-2}
\bibliography{main}

\end{document}